\documentclass[aps,prb,reprint,superscriptaddress,longbibliography,floatfix]{revtex4-1}
\usepackage{amsmath,amsfonts,amssymb,bm}
\usepackage[final]{graphicx}
\usepackage{float}
\usepackage[caption = false]{subfig}
\usepackage{dcolumn}
\usepackage[thinspace,mediumqspace,textstyle,amssymb]{SIunits}
\usepackage{epsfig}
\usepackage{color}
\usepackage{multirow}
\usepackage{epstopdf}
\usepackage[acronym]{glossaries}
\usepackage[EULERGREEK]{sansmath}
\usepackage[normalem]{ulem}
\usepackage[colorinlistoftodos,prependcaption,textsize=tiny]{todonotes}
\usepackage{color,soul}
\usepackage{tabularx}
\usepackage{array}
\usepackage{listings}
\usepackage{setspace}
\usepackage[normalem]{ulem}
\usepackage[toc,page]{appendix}

\begin{document}

\title[First-principles modeling of topological insulator surfaces]{First Principles Modeling of Topological Insulators: Structural Optimization and Exchange Correlation Functionals}

\author{Thomas~K.~\surname{Reid}}
\affiliation{Department of Materials Science and Engineering and Institute of Materials Science, University of Connecticut, Storrs, Connecticut 06269, USA}
\author{S.~Pamir~\surname{Alpay}}
\affiliation{Department of Materials Science and Engineering and Institute of Materials Science, University of Connecticut, Storrs, Connecticut 06269, USA}
\affiliation{Department of Physics, University of Connecticut, Storrs, CT 06269, USA}
\author{Alexander~V.~\surname{Balatsky}}
\affiliation{NORDITA, KTH Royal Institute of Technology and Stockholm University, Roslagstullsbacken 23, Stockholm SE-106 91, Sweden}
\affiliation{Department of Physics, University of Connecticut, Storrs, CT 06269, USA}
\author{Sanjeev~K.~\surname{Nayak}}
\email{sanjeev.nayak@uconn.edu}
\affiliation{Department of Materials Science and Engineering and Institute of Materials Science, University of Connecticut, Storrs, Connecticut 06269, USA}


\begin{abstract}
Topological insulators (TIs) are materials that are insulating in the bulk but have zero band gap surface states with linear dispersion and are protected by time reversal symmetry. These unique characteristics could pave the way for many promising applications that include spintronic devices and quantum computations. It is important to understand and theoretically describe TIs as accurately as possible in order to predict properties. Quantum mechanical approaches, specifically first principles density functional theory (DFT) based methods, have been used extensively to model electronic properties of TIs. Here, we provide a comprehensive assessment of a variety of DFT formalisms and how these capture the electronic structure of TIs. We concentrate on Bi$_2$Se$_3$ and Bi$_2$Te$_3$ as examples of prototypical TI materials. We find that the generalized gradient (GGA) and kinetic density functional (metaGGA) produce displacements increasing the thickness of the TI slab, whereas we see an opposite behavior in DFT computations using LDA. Accounting for van der Waals (vdW) interactions overcomes the apparent over-relaxations and retraces the atomic positions towards the bulk. Based on an intensive computational study, we show that GGA with vdW treatment is the most appropriate method for structural optimization. Electronic structures derived from GGA or metaGGA employing experimental lattice parameters are also acceptable. In this regard, we express a slight preference for metaGGA in terms of accuracy, but an overall preference for GGA due to compensatory improvements in computability in capturing TI behavior. 

\end{abstract}

\maketitle

\section{Introduction}

Within the last decade, it was reported that the pnictogen chalcogenides Bi$_2$Se$_3$ and Bi$_2$Te$_3$ exhibit the properties of a three-dimensional topological insulator (TI)---small energy gap at the Fermi level, inverted parity of the band edge states leading to a Dirac band dispersion, and quantum oscillations for topological surface states (TSS) ~\cite{Zhang-NP-2009,Xia-NP-2009,Hsieh-Nature-2009,Hasan-RMP-2010,Moore-Nature-2010,Zhang-NP-2010,Taskin-AM-2012,Tsipas-ACSN-2014}. This discovery has resulted in significant interest in these two materials, as their relatively large bulk bandgaps and chemical simplicity make the study of their topological electronic physics readily accessible to theory and experiment work alike. TIs and TSS are projected to be the basis of many cutting-edge device applications, including spintronics and quantum computers~\cite{Das-Sarma-PhyToday-2006,Wang-2016,Fan-Spin-2016}. TIs also offer fertile ground for fundamental studies of exotic electronic phenomena arising from wave function topology.~\cite{Hasan-RMP-2010,Zhang-PRL-2013-A,Zhang-PRL-2013-B,Kane-PhysScripta-2015,Metlitski-PRB-2015,Teo-PRB-2008,Hsieh-Science-2009,Liang-PRL-2008,Kane-PRL-2005,Wehling-AP-2014} 

Structurally, Bi$_2$Se$_3$ and Bi$_2$Te$_3$ are most often described in terms of quintuple layers (QLs), in which atomic layers are arranged in sets of five sublayers along the out-of-plane $c$-axis of Bi$_2$Se$_3$ and Bi$_2$Te$_3$, a configuration called the quintuple layer (QL). The arrangement of atoms in a QL is B(1)-A-B(2)-A-B(1) (for A$_2$B$_3$) along the [0001]-direction of the hexagonal crystal system with space group $\text{R}\bar{3}\text{m}$ (No. 166)~\cite{Nakajima-JPCS-1963}. The layers in a QL are bounded by stronger chemical bonding, but the inter-QL interaction is comparatively weaker and of van der Waals (vdW) type. 

The number of QLs needed to achieve a robust TSS is of primary interest for the accurate description of the TI behavior. Different factors such as quantum effects due to finite size, undercoordinated surface atoms, and hybridization of orbitals from the surface and bulk atoms also contribute to the emergence or suppression of a TSS. Zhang \textit{et al.}~\cite{Zhang-NP-2010} have reported the appearance of a Dirac point in samples with thicknesses 6QLs and greater for Bi$_2$Se$_2$, while, for Bi$_2$Te$_3$, Liu \textit{et al.}~\cite{Liu-PRB-2012} have reported that 4QLs are sufficient. Using a combined model Hamiltonian study and first-principles calculations Liu \textit{et al.}~\cite{Chao-Xing-PRB-2010} have reported an oscillatory crossover from a two-dimensional to a three-dimensional TI as a function of film thickness (number of QLs). Since both intra-QL and inter-QL interactions are crucial to stabilize the TSS~\cite{Young-PRB-2011,Liu-PRB-2011}, it is important to standardize the settings of density functional theory (DFT) modeling to converge towards a common perspective as applied to this general class of materials.  

DFT is one of the most popular theoretical tools to study structural and electronic properties of materials. It relies upon the Kohn-Sham formulation, where the knowledge of electron density is sufficient to calculate several materials properties that depend on the electronic and atomic structure. It is an efficient means of establishing the ground state, even for relatively large systems. Furthermore, because it is a first-principles method, it allows an unambiguous comparison to model experiments. It is important to note that DFT's robustness is not based only in its capacity for verification and explanation of experimental results. It also allows for predicting properties for previously unexplored systems to guide experimental work. While the validity of principles of DFT is established, the deficiencies of exchange-correlation functionals (XCFs) may limit the quality of its predictive capabilities. A substantial amount of knowledge has been accumulated to quantify these shortcomings, and ways to circumvent them have been suggested with mixed success of transferability~\cite{Cohen-Science-2008,Cohen-ChemRev-2012,Santra-JCP-2019,Zunger-Physics-2010,Adeagbo-PRB-2014,Nayak-PRB-2015,Dion-PRL-2004,Berland-PRB-2019,Liechtenstein-PRB-1995,Becke-JCP-2006,Zhao-JCP-2006}.     

For TI materials like Bi$_2$Se$_3$ and Bi$_2$Te$_3$, DFT can be utilized in two major ways: The first is the description of the ground-state atomic positions (relaxation). The second is the description of the converged charge density. Both play a significant role in quantifying TSS and the selection of the appropriate XCFs. TI materials have been studied with XCFs including the local density approximation (LDA) and the generalized gradient approximation (GGA)~\cite{Zhang-NP-2009,Zhang-NP-2010,Yazyev-PRB-2012,Yazyev-PRL-2010,Li-PRB-2014,Tao-NP-2018,Datzer-PRB-2017,Xia-NP-2009,Seixas-NatureComm-2015,Seibel-PRB-2016,Rusinov-JExptTheoPhys-2013}, but the role of the kinetic density functional (metaGGA) has not been hitherto explored. The many-body perturbation theory in the $GW$ approximation, where $G$ is the single particle Green function and $W$ is the screened Coulomb interaction, has been applied to improve description of the band structure~\cite{Aguilera-PRB-2013,Forster-PRB-2015,Forster-PRB-2016}. It is also important to realize that atomic structure is intricately related to electronic properties. For example, the $GW$ approximations are often applied to experimental or DFT-derived atomic structures. For layered TI materials like Bi$_2$Se$_3$ and Bi$_2$Te$_3$, treatment of the vdW forces has proven particularly important, and many studies have demonstrated that doing so greatly affects DFT's description of electronic structures. This is because, for a slab or film of a fixed number number of QLs, a relationship exists between the thickness of the slab or film (i.e. due to strain) and the possibility of observing a TSS. VdW forces mediate the interactions between QLs across the inter-QL space---depending on the magnitude of their influence, they create smaller or larger vdW gaps. As a result, an XCF treated to explicitly account for the vdW forces might predict a different overall thickness than one that has not been, and, thus, it might predict a different surface state~\cite{Liu-PRB-2013,Luo-PRB-2012}. However, it must be noted that the vast majority of the thickness of a slab is made up of the QLs themselves. In order to satisfactorily account for a given XCF's contribution to change in thickness relative to the experimental parameters and those produced by other XCFs, we must assess the role played by the intra-QL space in overall thickness as well, which has not been taken into account by previous studies.

In this study, we carry out a systematic analysis in pursuit of a consensus regarding the role of XCFs in the outcome of structural optimization and its subsequent influence over electronic structure calculations in Bi$_2$Se$_3$ and Bi$_2$Te$_3$. We specifically asses a variety of XCFs including: LDA, GGA, metaGGA, GGA+vdW, and metaGGA+vdW. We find that the structural optimization process is highly sensitive to the type of XCF employed. LDA and GGA produce relaxation trends opposite of each other. The effective role of the vdW treatment is to bring the relaxation toward the experimental bulk positions. While the outcomes of the electronic structure calculations are dependent on the specific XCF, we used the vdW treatments of Grimme \textit{et al.}~\cite{Grimme-JCP-2010}, which do not depend on charge density, making it redundant to apply them to the self-consistent charge density (SCCD) and band structure phases of our calculations. As a result, our vdW-treated XCFs preserve the charge density distribution generated by their root functionals. It is important to note that our structural optimizations with the vdW treatments indicate little deviation (below 2\%) from experimental bulk coordinates of Nakajima (Ref.~\cite{Nakajima-JPCS-1963}). Therefore, we deduce that the experimental bulk coordinates of Nakajima are sufficient to create structural models of Bi$_2$Se$_3$ and Bi$_2$Te$_3$ in which TSSs that exhibit thickness-dependence in agreement with experiment can be expected, so long as the GGA or metaGGA XCFs (the root functionals of our vdW-corrected XCFs) are used. Lastly, mindful of the range of XCFs, number of QLs (slab thicknesses), and computational strategies pursued in the theoretical literature on TSSs of Bi$_2$Se$_3$ and Bi$_2$Te$_3$, we extend these conclusions to a prescription for a general method for generating models suitable for studying TI physics in binary layered systems of space group $R\bar{3}m$ (No. 166), with DFT.

In this way, we seek to evaluate with new precision the outcome of structural optimizations with a variety of XCFs and benchmark a procedure for reliably modeling TSSs in these systems. We expect this study should act as the groundwork for the production of models that might be used to generate data that would serve the development of novel devices.

\section{Methodology}

\begin{figure*}[t]
\subfloat{\includegraphics[width = 0.85\linewidth]{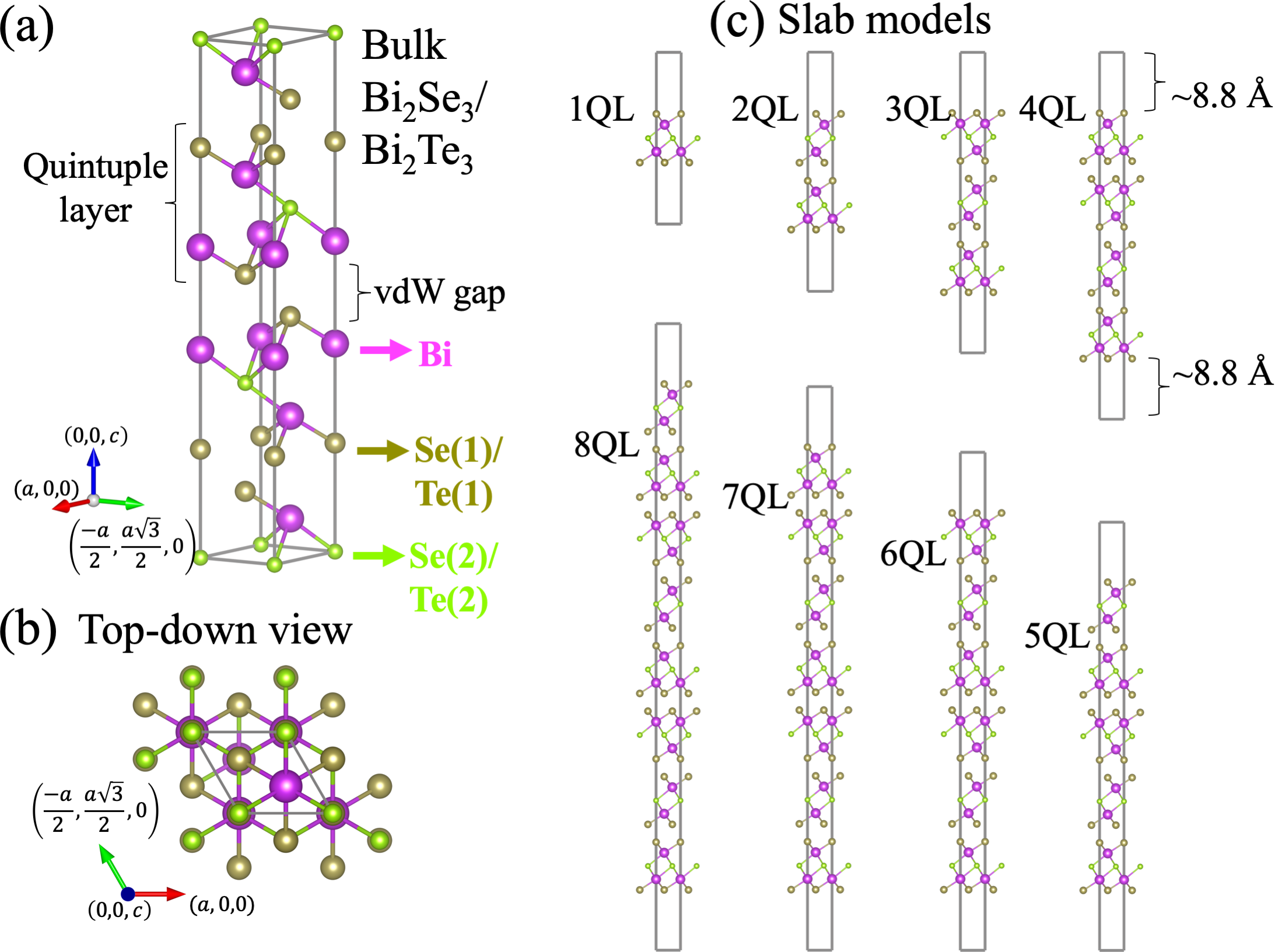}} 
\caption{Schematic pictures of bulk crystal lattice of Bi$_2$Se$_3$ and Bi$_2$Te$_3$ in hexagonal system with space group $R\bar{3}m$ (166) in (a) and (b). Slab models constructed from the bulk lattice with varied quintuple layers are shown in (a). }
\label{figure1}
\end{figure*}

Calculations were performed with DFT using the Vienna \textit{Ab initio} Simulation Package (VASP)~\cite{Kresse-CMS-1996,Kresse-PRB-1996} which solves the scalar-relativistic Kohn-Sham Hamiltonian. In all calculations, spin-orbit coupling was treated with the perturbation theory. We used the following three root XCFs: LDA with Ceperley-Alder parametrization~\cite{Ceperley-PRL-1980}, GGA with Perdew-Burke-Ernzerhof parametrization (PBE)~\cite{Perdew-PRL-1996}, and metaGGA with Strongly Constrained and Appropriately Normed (SCAN) approach~\cite{Sun-PRL-2015}. Slab models were prepared from the unit bulk hexagonal lattice with lattice parameters taken from experiments~\cite{Nakajima-JPCS-1963}. This set of experimental lattice parameters matches the DFT band structure as reported in Ref.~\cite{Luo-PRB-2012}. The schematic picture of the bulk hexagonal unitcell and the slab models are shown in Fig.~\ref{figure1}(a--c). The slabs were centered along the vertical direction. A vacuum layer of thickness of $\sim$8.8~{\AA} was added to either side of the surfaces making a total vacuum thickness of $\sim$17.6~{\AA}. This range of vacuum thickness ensures negligible dipole interaction between the surfaces through the vacuum and is regularly used for modeling surfaces and nanomaterials~\cite{Sahoo-SurfaceScience-2018,Sahoo-PRB-2010}. A systematic study from one quintuple layer (1QL) to eight quintuple layers (8QL) models were performed. The calculations were performed using the projector-augmented wave pseudopotentials, a symmetrized 7$\times$7$\times$1 $k$-point grid to span the two-dimensional Brillouin zone, the kinetic energy cutoff of the plane-waves set to 500 eV and the precision tag set to `accurate'. An additional support grid was included in the calculation through the tag `addgrid', which accounts for accurate inter-atomic forces and hence leads to higher quality geometrical optimizations. The total energy convergence for self-consistent field cycles is set to $10^{-7}$~eV and the force convergence for geometrical optimization is set to $0.001$~eV/{\AA}.

Each QL of a multi-QL system is weakly bounded to the others by the vdW force. Past works have demonstrated the importance of accounting for the influence of the vdW forces in layered systems like Bi$_2$Se$_3$ and Bi$_2$Te$_3$ in DFT calculations by using specially adapted treatments~\cite{Luo-PRB-2012, Liu-PRB-2013}. We incorporated such treatments into another set of calculations with the GGA functional using the zero-damping DFT-D3 method of Grimme {\it et al.}~\cite{Grimme-JCP-2010}. Although the metaGGA-SCAN functional accounts for vdW interactions to some extent, we wanted to cross-check the outputs including the vdW interactions, so a set of calculations using the SCAN+vdW treatment was carried out.

\section{Structural Properties}

The experimental bulk lattice parameters and corresponding results from the DFT calculations for Bi$_2$Se$_3$ and  Bi$_2$Te$_3$ are shown in Tab.~\ref{bulk_LC_BiSe} and Tab.~\ref{bulk_LC_BiTe}, respectively. Our calculations are delineated by rows by XCF. The data shows that SCAN+vdW predictions of lattice parameters most closely reflect experimental results, followed in this regard by PBE+vdW. LDA and PBE XCFs deviate more from the experimental values, with LDA predicting structural parameters lower than experiment and PBE predicting structural parameters higher than experiment. Relaxation along $a$ resulted in far less change overall compared with relaxations along $c$.

\begin{table}[h]
\caption{\label{bulk_LC_BiSe}Table showing the lattice contants $a$ and $c$, Wyckoff reference of Bi, Se(1) and Se(2) positions, and optical bandgap of bulk Bi$_2$Se$_3$ obtained from various treatments of XCFs in DFT and the corresponding experimental values.} 
\begin{ruledtabular}
\begin{tabular}{cccccc}
  $a$ ({\AA}) & $c$ ({\AA}) & $z$-Bi & $z$-Se(1) & $z$-Se(2) & E$_{\text{g}}$ (eV)   \\
  \hline
  4.108$^\text{a}$ & 27.293$^\text{a}$ & 0.4012$^\text{a}$ & 0.0$^\text{a}$ & 0.2093$^\text{a}$ & 0.042 (0.326)$^\text{a}$  \\
  4.188$^\text{b}$ & 31.429$^\text{b}$ & 0.3953$^\text{b}$ & 0.0$^\text{b}$ & 0.2216$^\text{b}$ & 0.691 (0.302)$^\text{b}$ \\
  4.174$^\text{c}$ & 28.872$^\text{c}$ & 0.4001$^\text{c}$ & 0.0$^\text{c}$ & 0.2125$^\text{c}$ & 0.234 (0.198)$^\text{c}$ \\
  4.157$^\text{d}$ & 29.812$^\text{d}$ & 0.3983$^\text{d}$ & 0.0$^\text{d}$ & 0.2160$^\text{d}$ & 0.353 (0.058)$^\text{d}$ \\
  4.140$^\text{e}$ & 28.657$^\text{e}$ & 0.4004$^\text{e}$ & 0.0$^\text{e}$ & 0.2121$^\text{e}$ & 0.115 (0.278)$^\text{e}$ \\
  4.143$^\text{f}$ & 28.636$^\text{f}$ & 0.4008$^\text{f}$ & 0.0$^\text{f}$ & 0.2117$^\text{f}$ & 0.220$^\text{g}$, 0.335$^\text{h}$ \\
\end{tabular}
\end{ruledtabular}
\begin{flushleft}
$^\text{a}$  LDA (+SOC), present calculation.\\
$^\text{b}$  PBE (+SOC), present calculation.\\
$^\text{c}$  PBE+vdW (+SOC), present calculation.\\
$^\text{d}$  SCAN (+SOC), present calculation.\\
$^\text{e}$  SCAN+vdW (+SOC), present calculation.\\
$^\text{f}$  Experiment Ref.~\cite{Nakajima-JPCS-1963}\\
$^\text{g}$  Experiment Ref.~\cite{Martinez-SciRep-2017}\\
$^\text{h}$  Experiment Ref.~\cite{Nechaev-PRB-2013}
\end{flushleft}
\end{table}

\begin{table}[h]
\caption{\label{bulk_LC_BiTe}Table showing the lattice contants $a$ and $c$, Wyckoff reference of Bi, Se(1) and Se(2) positions, and optical bandgap of bulk Bi$_2$Te$_3$ obtained from various treatments of XCFs in DFT and the corresponding experimental values. } 
\begin{ruledtabular}
\begin{tabular}{cccccc}
  $a$ ({\AA}) & $c$ ({\AA}) & $z$-Bi & $z$-Se(1) & $z$-Se(2) & E$_{\text{g}}$ (eV)   \\
  \hline
  4.356$^\text{a}$ & 29.903$^\text{a}$ & 0.4016$^\text{a}$ & 0.0$^\text{a}$ & 0.2089$^\text{a}$ & 0.138 (0.130)$^\text{a}$  \\
  4.445$^\text{b}$ & 32.247$^\text{b}$ & 0.3970$^\text{b}$ & 0.0$^\text{b}$ & 0.2163$^\text{b}$ & 0.555 (0.031)$^\text{b}$ \\
  4.432$^\text{c}$ & 30.446$^\text{c}$ & 0.3998$^\text{c}$ & 0.0$^\text{c}$ & 0.2105$^\text{c}$ & 0.198 (0.244)$^\text{c}$ \\
  4.411$^\text{d}$ & 31.474$^\text{d}$ & 0.3981$^\text{d}$ & 0.0$^\text{d}$ & 0.2139$^\text{d}$ & 0.391 (0.079)$^\text{d}$ \\
  4.400$^\text{e}$ & 30.096$^\text{e}$ & 0.4001$^\text{e}$ & 0.0$^\text{e}$ & 0.2096$^\text{e}$ & 0.179 (0.169)$^\text{e}$ \\
  4.386$^\text{f}$ & 30.497$^\text{f}$ & 0.4000$^\text{f}$ & 0.0$^\text{f}$ & 0.2097$^\text{f}$ & 0.171$^\text{g}$ \\
\end{tabular}
\end{ruledtabular}
\begin{flushleft}
$^\text{a}$  LDA (+SOC), present calculation.\\
$^\text{b}$  PBE (+SOC), present calculation.\\
$^\text{c}$  PBE+vdW (+SOC), present calculation.\\
$^\text{d}$  SCAN (+SOC), present calculation.\\
$^\text{e}$  SCAN+vdW (+SOC), present calculation.\\
$^\text{f}$  Experiment Ref.~\cite{Nakajima-JPCS-1963}\\
$^\text{g}$ Experiment Ref.~\cite{Che-Yu-JAP-1961}
\end{flushleft}
\end{table}

\begin{figure*}[!]
\includegraphics[width = 0.9\linewidth]{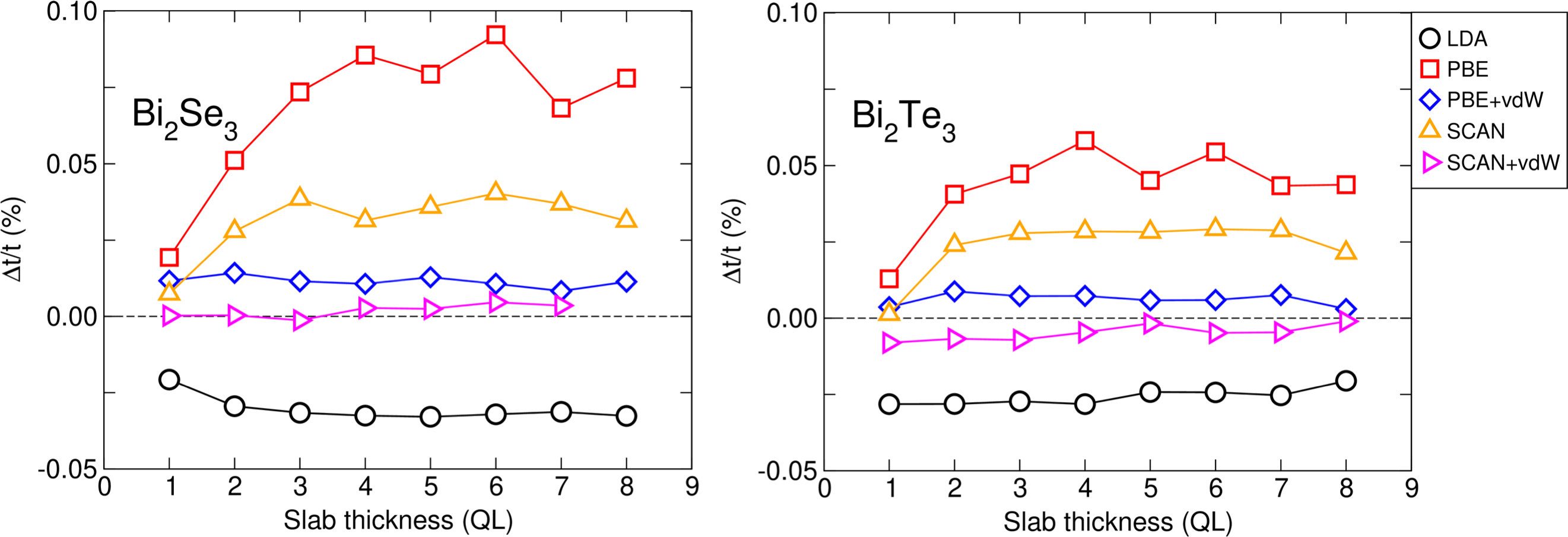}
\caption{Change in slab thickness expressed in percent with respect to bulk values obtained after geometrical optimization using different XCFs for (a) Bi$_2$Se$_3$ and (b) Bi$_2$Te$_3$.}
\label{figure2}
\end{figure*}

\begin{figure*}[!]
\includegraphics[width=0.9\linewidth]{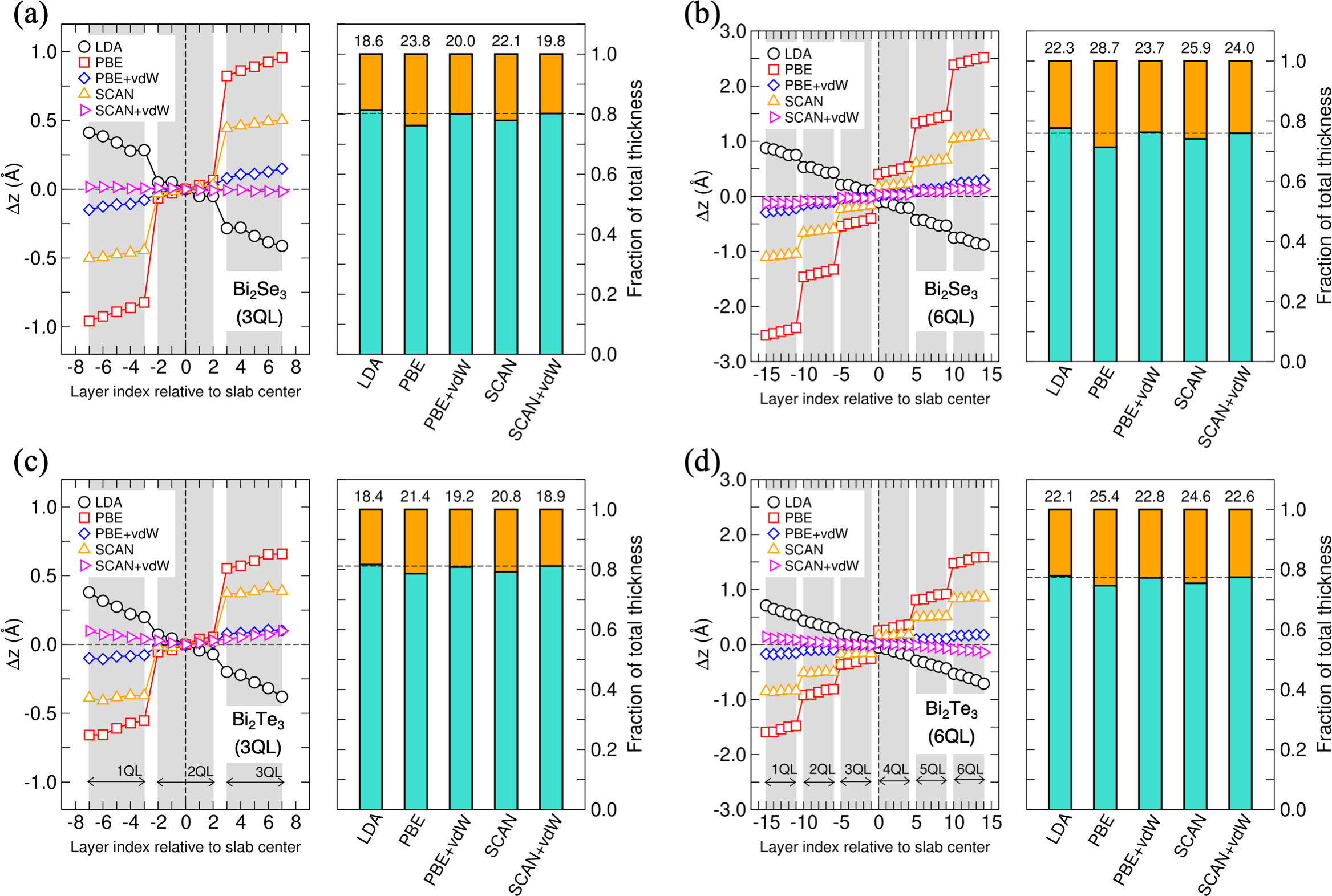}
\caption{Relaxation of layers in 3QL and 6QL slabs for various treatment of XCFs for Bi$_2$Se$_3$ ((a) and (b)) and Bi$_2$Te$_3$ ((c) and (d)). Left panel is the difference between relaxed and bulk atomic positions $\Delta z$. The zero value of layer index (abscissa) is mid way between the slab thickness~(see ~\ref{figure1}). The right panel is the fraction of inter-QL separation (orange color) and QL thickness (turquoise color) contribution to the total thickness. The fraction of inter-QL separation in percentage is displayed in numbers.} 
\label{figure3}
\end{figure*}

\begin{figure}[!]
\includegraphics[width=0.8\linewidth]{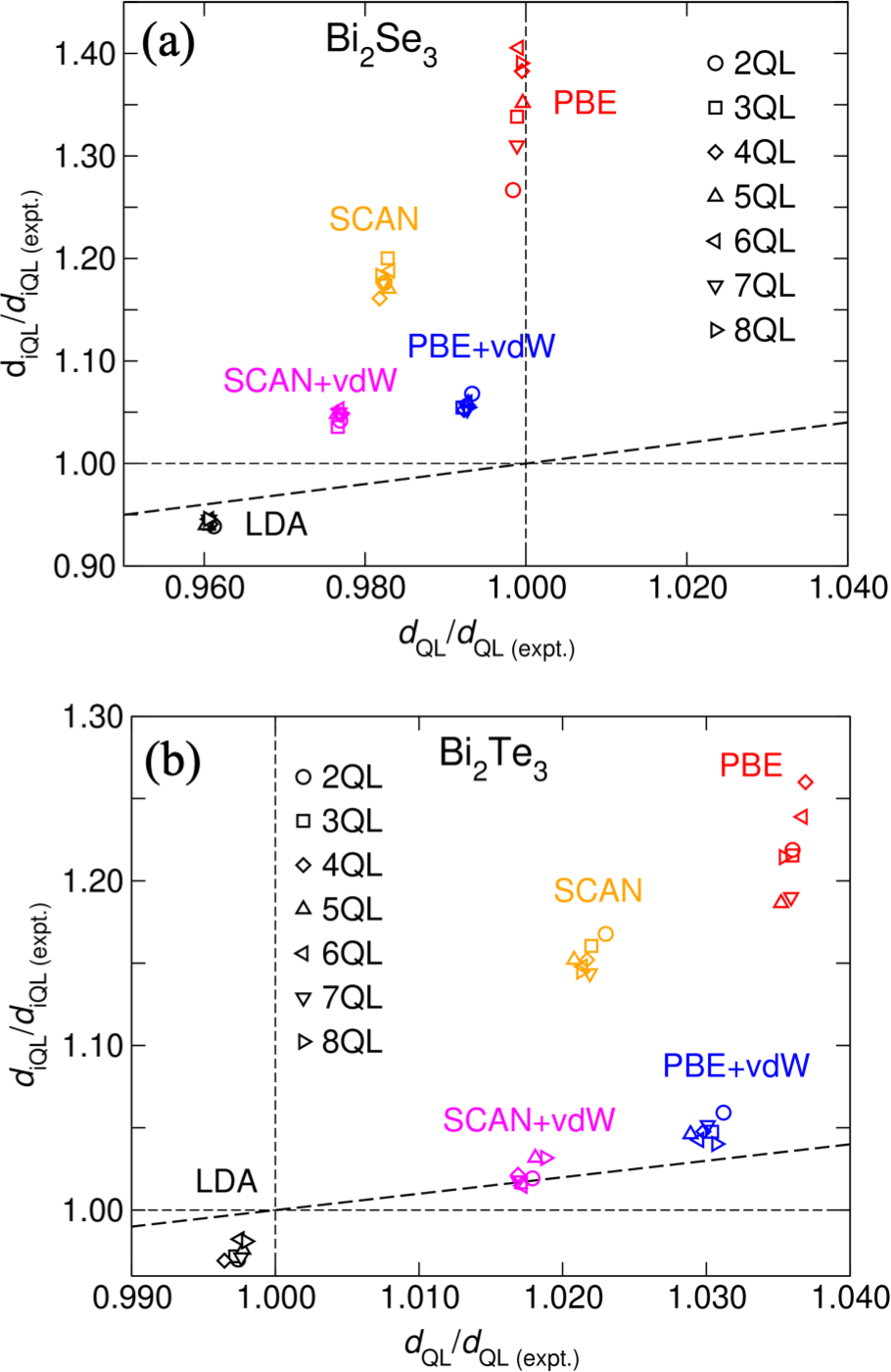}
\caption{Ratio of the optimized outer inter-QL separation ($d_{\text{QL}}$) and top inter-QL separation ($d_{\text{iQL}}$) obtained from different XCFs for 2--8QL models to their respective ultrathin film experimental values. The experimental values are taken from Ref.~\cite{Roy-PRB-2014} for Bi$_2$Se$_3$ and  Ref.~\cite{Fukui-PRB-2012} for Bi$_2$Te$_3$.} 
\label{intra-inter-QL}
\end{figure}

\begin{figure*}[t]
\includegraphics[width = .8\linewidth]{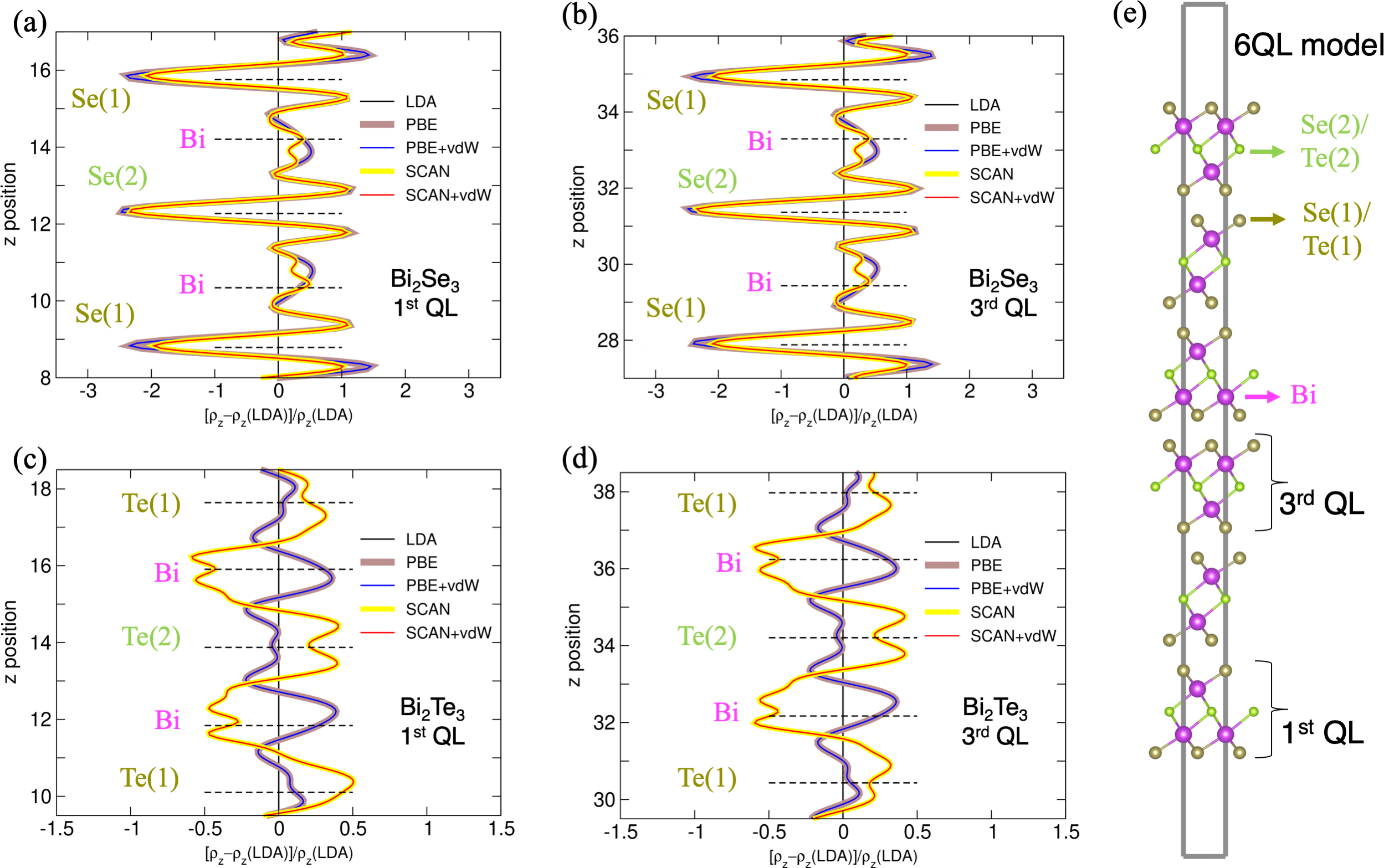}
\caption{A comparison of charge density obtained from SCF calculations of various XCFs using a 6QL slab model with atomic positions of bulk experimental coordinates. The first QL (surface QL) and third QL (inner QL) of the slab model is shown for Bi$_2$Se$_3$ in (a) and (b), respectively, and similarly for Bi$_2$Te$_3$ in (c) and (d). The intra-atomic positions of the QLs are marked with horizontal dashed lines.  There are notable differences on the charge density distribution for different XCFs and vdW corrected DFT preserving the the charge density of the corresponding root functional (PBE or SCAN). }
\label{figure5}
\end{figure*}
The slab models generated from the bulk experimental coordinates of Ref.~\cite{Nakajima-JPCS-1963} show variation in the direction and extent of relaxation with different XCF treatments. The relative change of thickness of the slab $\Delta t$ = $t_{\text{XCF}}-t_{\text{Expt.}}$, where $t_{\text{XCF}}$ and $t_{\text{Expt.}}$ are the optimized thickness and ideal bulk-cut thickness, respectively, as a function of number of QLs for Bi$_2$Se$_3$ and Bi$_2$Te$_3$ is shown in Fig.~\ref{figure2}(a) and Fig.~\ref{figure2}(b), respectively. The horizontal line passing through zero for $\Delta t$ represents the thickness of the structure with experimental lattice parameters. For all calculations conducted with the LDA functional, a negative value of $\Delta t$ is observed. Naturally, this implies the LDA functional induces a reduction in the thickness of all slabs. An opposite trend is observed for the PBE functional, implying that PBE predicts the thickness of any slab should increase. The SCAN functional, for both materials, predicts a less severe increase of the slab thickness than PBE. When we apply the vdW treatments to both PBE and SCAN, both functionals predict that the thickness of the slab should be bulk-like. These trends, on a functional-by-functional basis, are reflected in the resultant bulk coordinates of each XCF (Tab.~\ref{bulk_LC_BiSe} and Tab.~\ref{bulk_LC_BiTe}). By introducing SOC, the bandgap shrinks. This is consistent with the expectation that these materials are composed of high-$Z$ elements that have sizable SOC interactions. It is notable, however, that, for the Bi$_2$Se$_3$ LDA model, enabling SOC increases the bandgap. This result is opposite those for all other bulk models of both Bi$_2$Se$_3$ and Bi$_2$Te$_3$. We surmise that this is due to the large reduction in the $a$ lattice parameter.

While it is obvious each functional affects the general thickness of the slab, a focus on the thickness alone obscures the control each XCF exerts over the role of more specific mechanisms in the outcome of a given structural optimization. In the left panels of Fig.~\ref{figure3}, we illustrate the change in the position of each atom along $c$ relative to its original index position for 3QL and 6QL models for Bi$_2$Se$_3$ (Fig.~\ref{figure3}(a) and (b)) and Bi$_2$Te$_3$ (Fig.~\ref{figure3} (c) and (d)), respectively, taking into account the five discussed XCFs. The $\Delta z$ (= $z_{\text{relaxed}}$ $-$ $z_{\text{initial}}$) is the amount of change incurred by an atom in comparison to its position in the initial model generated from the experimental bulk lattice parameters. Since the zeroth index in the left panel of Figs.~\ref{figure3}(a)--(d) is set to the center layer of the slab for models with an odd number of QLs and one of the two center layers for models with an even number of QLs, an atom's position relative to the horizontal at zero is indicative of the direction of its relaxation. Presented such a way, the positive value of $\Delta z$ for the negative layer indices and negative value of $\Delta z$ for the positive layer indices imply inward relaxation of the two ends of the slab. Similarly, the negative value of $\Delta z$ for the negative layer indices and positive value of $\Delta z$ for the positive layer indices imply outward relaxation. The trend of the change in the layer positions within a QL---identifiable as the five-point clusters demarcated by the grey boxes---is generally linear, inward or outward, for all calculations. A step between each QL represents the vdW gap (not to be confused with the vdW treatments for XCFs). PBE predicts the largest outward change in the size of the vdW gap relative to the experimental observations. PBE+vdW and SCAN+vdW, on the other hand, predict a far smaller change in the vdW gap, slightly inward or outward depending on the system and the thickness of the slab (refer to the left panel). The remaining functionals predict absolute changes in the vdW gap larger than the vdW-corrected functionals, but smaller than pure PBE. Comparing to the bulk, which is the horizontal line passing through $\Delta z = 0$ in the left panel of Figs.~\ref{figure3}(a)--(d), it is observed that SCAN+vdW and PBE+vdW deviate minimally from the corresponding bulk atomic positions. Roy~\textit{et al.}~\cite{Roy-PRB-2014,Roy-PRL-2014} have performed surface X-ray diffraction studies on (0001) Bi$_2$Se$_3$ grown as an ultrathin film on Si(111) using MBE. They find outward relaxation of the top Se-Bi layer by $\sim$2\%-4\% as compared to the bulk coordinates, and $\sim$3\% contraction of the top inter-QL separation compared to the same. Our results suggest that PBE+vdW relaxation match more closely to the experimental results for the outer Se-Bi layers. However, the contraction of the top inter-QL layer is not represented in our calculations. 

It is clear from the right panel of Figs.~\ref{figure3}(a)--(d) that the majority of the thickness of each relaxed slab is occupied by the intra-QL space and a minority by the inter-QL space. That the thickness of the vdW-corrected functionals maintains approximately the same ratio of intra-QL space to inter-QL space as the other functionals, but exhibits far lower thickness in general, suggests two possibilities: 1. By merit of the large absolute differences between the thicknesses of the non-vdW-corrected and vdW-corrected models, that the vdW force is the primary mediating force during structural optimization. 2. That the vdW force controls the inter-QL distances as well as the the intra-QL distances.

For LDA, PBE, and SCAN models, even when they are as large as 8QL, convergence to the experimental bulk characteristics is not obtained. On the other hand, it should again be noted that the relaxations of the vdW-corrected models, with a slight bias in favor of the SCAN+vdW models in particular, show strong convergence to the corresponding experimental bulk for any number of QLs. This suggests that slabs prepared from the experimental coordinates, and not subject to structural optimization, adequately capture the physics accounted for by the vdW treatments, and are reasonable for studying the electronic structure. But, a conclusion can not be drawn without evaluating the electronic structure directly. This is done in the next section.

In Fig.~\ref{intra-inter-QL}, we plot the relationship between the ratio of the optimized inter-QL space (the most outward intra-QL), $d_{\text{QL}}$, and inter-QL space (the top Bi-Te or Bi-Se layer separation), $d_{\text{iQL}}$, to their respective ultrathin film experimental values. The experimental values are taken from Ref.~\cite{Roy-PRB-2014} for Bi$_2$Se$_3$ and  Ref.~\cite{Fukui-PRB-2012} for Bi$_2$Te$_3$. In both panels, the horizontal and vertical dashed lines intersect at the experimental data point, and the diagonal dashed lines represent an extrapolation of the ratio of the portions of the total thickness occupied by the inter-QL space to the intra-QL space in the experimental films for slabs of variable thickness. The advantage of this set of plots is that the contributions of the changes in the $d_{\text{iQL}}$ and the $d_{\text{QL}}$  to the overall change in thickness after relaxation can be accounted for explicitly. For Bi$_2$Se$_3$, PBE induces a large change in the $d_{\text{iQL}}$; enabling the vdW treatment greatly reduces the predicted growth in the $d_{\text{iQL}}$, while also inducing a comparable reduction in the $d_{\text{QL}}$. By contrast, for Bi$_2$Te$_3$, PBE greatly increases both $d_{\text{QL}}$  and $d_{\text{iQL}}$ inter-QL separation, and in a fashion that deviates strongly from the experimental ratio (the dashed diagonal line). Enabling the vdW treatment leads to a convergence of the data to the experimental ratio, yet not to the experimental values reported in Ref.~\cite{Fukui-PRB-2012}. This again implies that the vdW treatment controls the overall thickness by changing both the $d_{\text{iQL}}$ and $d_{\text{QL}}$. For both of the materials, a similar trend can be seen with the SCAN functional and its vdW-treated counterpart. For both materials, LDA predicts a smaller $d_{\text{iQL}}$ and $d_{\text{QL}}$.

\section{Electronic Properties}
In this section, we seek to relate the selection of the XCF to the relationships between sample thickness and the TSS observed in experiment using the optimized structures generated in the previous section. 

\begin{figure}[h]
\subfloat{\includegraphics[width =.75\linewidth]{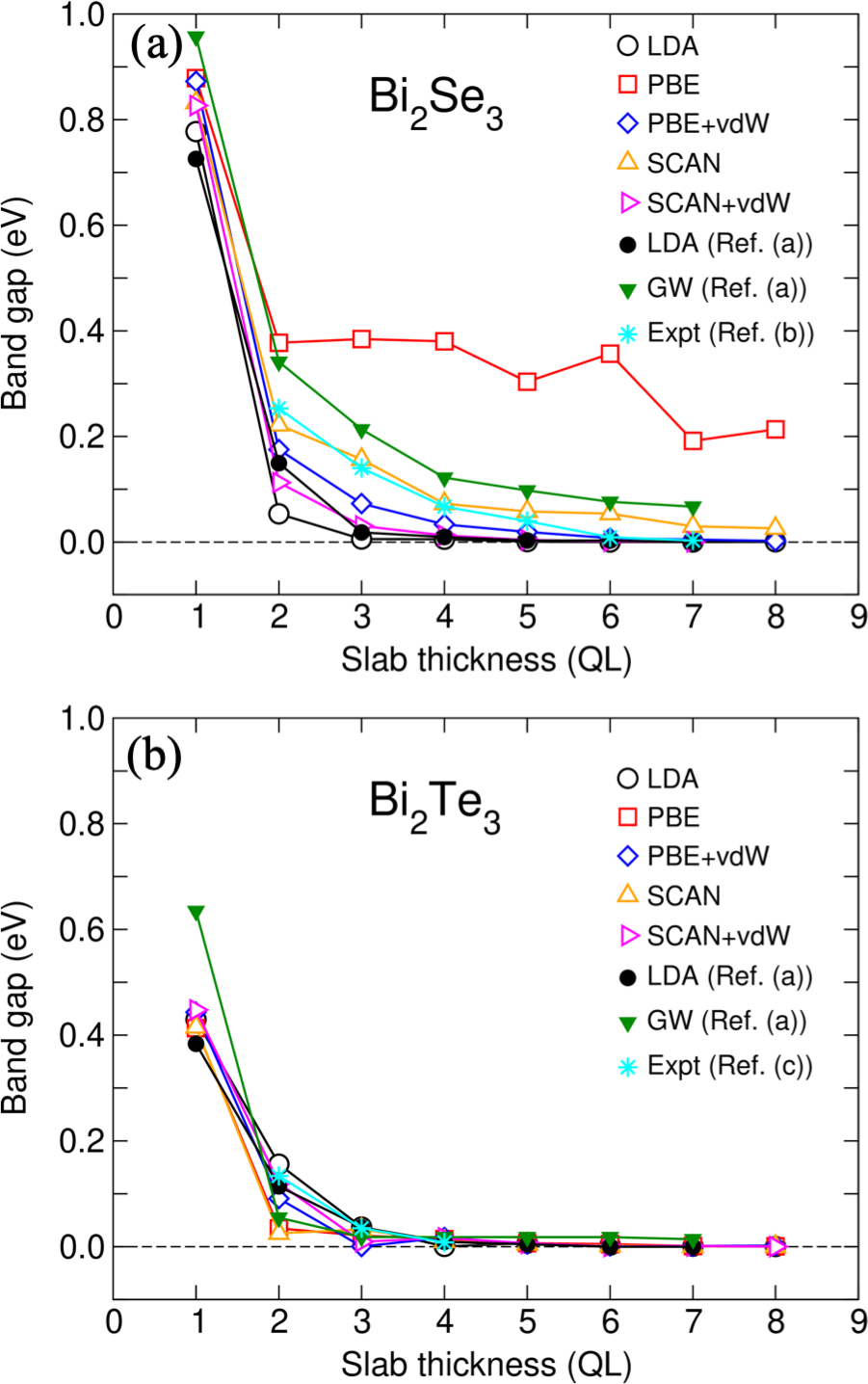}}
\caption{Band gap as a function of number of QL for Bi$_2$Se$_3$ and Bi$_2$Te$_3$ shown in (a) and (b), respectively. Some data is compared to literature with (a), (b) and (c) for data source referring to Ref.~\cite{Yazyev-PRB-2012}, Ref.~\cite{Zhang-NP-2010}, and Ref.\cite{Liu-PRB-2012}, respectively.}
\label{Figure5}
\end{figure}

\begin{figure}
\subfloat{\includegraphics[width = .9\linewidth]{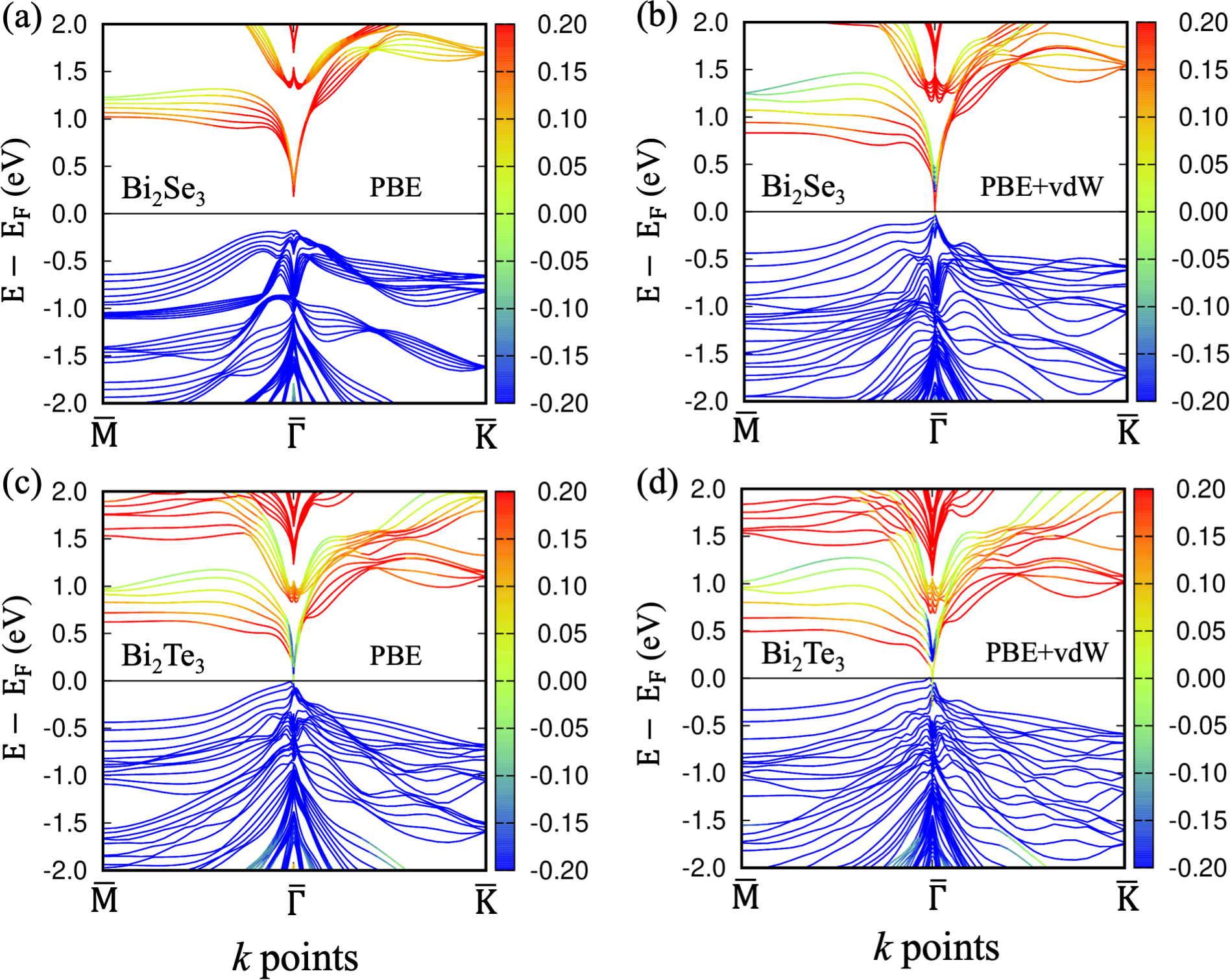}}
\caption{Band structure of 6QL Bi$_2$Se$_3$ ((a) and (b)) and 6QL Bi$_2$Te$_3$ ((c) and (d)) compared for PBE (left) and PBE+vdW (right). The band structure is colored according to relative weight of atomic occupancy, blue and red signifying larger contribution of Se/Te and Bi, respectively (see Fig.~\ref{figure7}). The bandgap in surface states is sensitive to the thickness of the slab and the relaxation effects, which is a consequence of XCFs chosen in the calculations. $\overline{\text{M}}$, $\overline{\Gamma}$, and $\overline{\text{K}}$ are the high-symmetry points of middle of the side, center, and vertex, respectively, of the surface Brillouin zone which is a two-dimensional hexagon.}
\label{figure6}
\end{figure}


First, however, a 6QL model generated from experimental bulk lattice parameters is chosen to show the $xy$-integrated charge density as a function of $z$. Further, the charge density is compared with the LDA-obtained charge density as the reference by the construct  $\left(\rho_z - \rho_z\text{(LDA)}\right)/\rho_z\text{(LDA)}$.  The comparison of charge density is done for the outer QL and one of the inner QLs (third QL from the outer QL) of Bi$_2$Se$_3$ in Fig.~\ref{figure5}(a) and  Fig.~\ref{figure5}(b), respectively, and similarly for Bi$_2$Te$_3$ in Fig.~\ref{figure5}(c) and Fig.~\ref{figure5}(d). The results demonstrate that the charge densities of PBE and SCAN show different trends as compared to LDA, and do not change with respect to where the QL is located relative to the center of the model. These 6QL unrelaxed models give band gap values of 0.0002~eV,  0.0027~eV,  0.0027~eV, 0.0050~eV, and 0.0050~eV for LDA, PBE, PBE+vdW, SCAN and SCAN+vdW XCFs for Bi$_2$Se$_3$. Similarly for Bi$_2$Te$_2$, the bandgaps are 0.0007~eV, 0.0012~eV, 0.0012~eV, 0.0008~eV, and 0.0008~eV for LDA, PBE, PBE+vdW, SCAN and SCAN+vdW, respectively. The data suggests that the bandgaps are controlled by the root functionals, with no change appearing with the addition of vdW treatments. 

Fig. \ref{Figure5} shows the evolution of the bandgap as a function of slab thickness for Bi$_2$Se$_3$ and Bi$_2$Te$_3$ with all XCFs. It is evident that the best agreement to experimental trends is achieved for Bi$_2$Se$_3$ by the SCAN functional, followed closely by the vdW-corrected functionals and the $GW$ method. For Bi$_2$Te$_3$, best agreement to experimental trends is achieved by the SCAN+vdW functional, followed closely by LDA and PBE+vdW. Given issues of computational efficiency and convergence we confronted with the SCAN functional, and the known tendency of LDA to underestimate bandgap, we conclude PBE+vdW strikes the best balance between computability and accuracy of all the functionals tested for both systems.

\begin{figure}
\subfloat{\includegraphics[width = .65\linewidth]{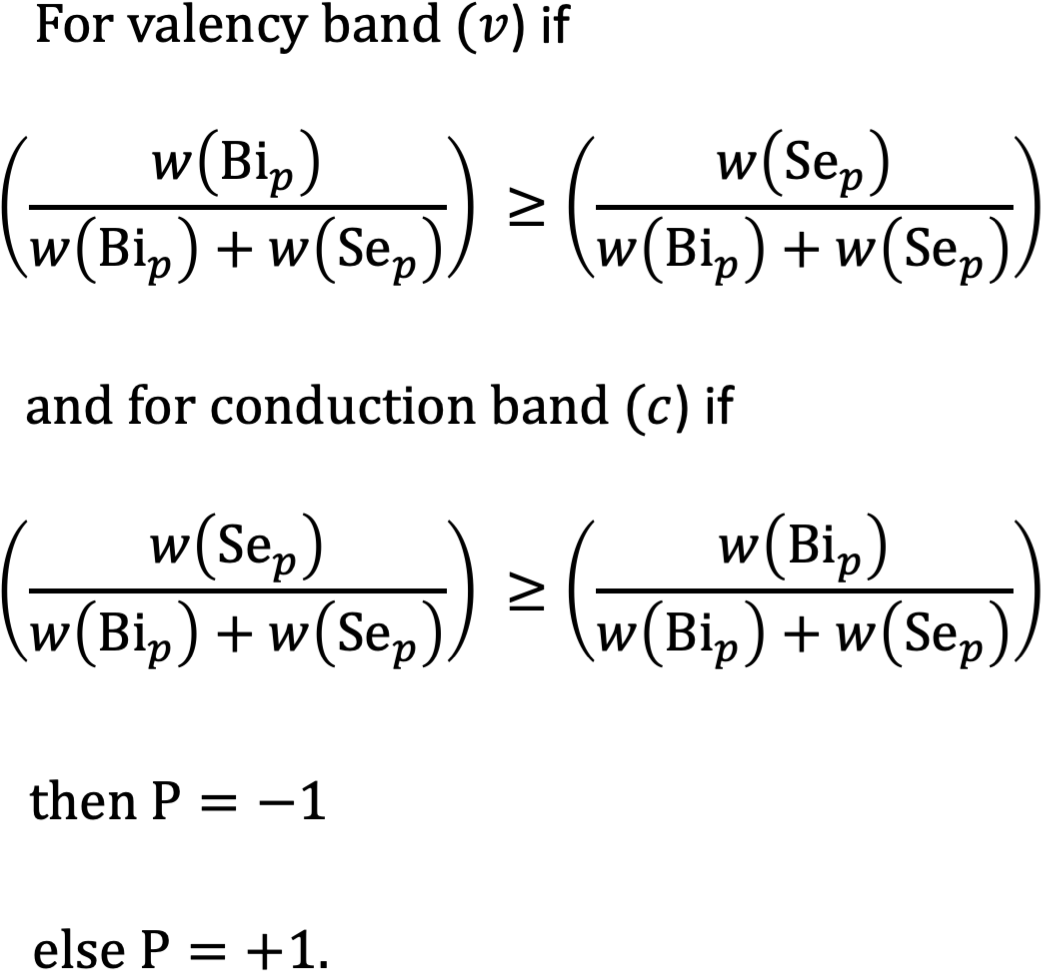}}
\caption{Algorithm used to compute the parity of states from the DFT band structure. Bi$_2$Se$_3$ is represented here as an example.}
\label{figure7}
\end{figure}

We now attempt to theoretically evaluate the TSSs of our models by examining other properties characteristic of the TSS, besides the bandgap: band parity, energy gradient with respect to $k$, and effective mass of the highest occupied band and lowest unoccupied band. The ideal TSS is characterized by linear band dispersion and a bandgap tending to zero. For $k$ close to $\Gamma$, the band energy can be expanded in $k$ using the Taylor's expansion.
\begin{equation}\label{TaylorExpansionEnergy}
E(k) = E_{0} + \left(\frac{\partial E}{\partial k}\right)k + \frac{1}{2} \left(\frac{\partial^2 E}{\partial k^2}\right)k^2 + O(k^3),
\end{equation}
where $E_{0}$ is band energy at the $\Gamma$-point. In the limit of the ideal TSS;
\begin{equation}
 \frac{\partial^2 E}{\partial k^2} \rightarrow 0.
\end{equation}
In real materials, we see a deviation from this ideal condition. Knowledge of $\left(\frac{\partial E}{\partial k}\right)$ and  $\left(\frac{\partial^2 E}{\partial k^2}\right)$ at the $\Gamma$-point provides an idea of the degree of deviation from the ideal TSS. The quantity $\left(\frac{\partial^2 E}{\partial k^2}\right)$ is an important property for carrier charges as it is related to the effective mass via the relation
\begin{equation}\label{EffectiveMass}
m^* = \frac{1}{\hbar^2}\left(\frac{\partial^2 E}{\partial k^2}\right).
\end{equation}
The calculated value of $\left(\frac{\partial E}{\partial k}\right)$ and $m^*$ in the direction $\Gamma\rightarrow\text{M}$ and $\Gamma\rightarrow\text{K}$ obtained from finite difference method is shown in Tab.~\ref{tab_em_BiSe} and Tab.~\ref{tab_em_BiTe}  for Bi$_2$Se$_3$ and Bi$_2$Te$_3$, respectively, for models varying in slab thickness (1QL--8QL) and XCFs. We find that the gradient $\partial E/\partial k$ of valence ($v$) and conduction bands ($c$) obtained from PBE have smaller values as compared to other XCFs for Bi$_2$Se$_3$. However, a similar tendency is not reflected for Bi$_2$Te$_3$. The effective masss of conduction band $m^{*}_{c}$ obtained from different XCF treatments are similar for Bi$_2$Se$_3$, while $m^{*}_{c}$ and $m^{*}_{v}$ have similar values (within reasonable error) for Bi$_2$Te$_3$.
The parity is calculated with the algorithm taken from Ref.~\cite{Geilhufe-PRB-2015} (see Fig.~\ref{figure7}, stated for Bi$_2$Se$_3$ as an example). The description of parity using this algorithm indicates the band inversion tendency for $P = -1$. Since the TSS is a limiting case, the parity of TSS can turn out to be $+1$ or $-1$ for such an algorithm.

Using a slightly different flavor of GGA (optPBE) and vdW treatments, the results in the band gap obtained for Bi$_2$Se$_3$ goes to zero smoothly as a function of thickness~\cite{Liu-PRB-2013}, which is similar to our results. However, the band gap of Bi$_2$Te$_3$ appears to converge towards zero from 2QL onwards, which in our case (PBE+vdW) goes to zero for 3QL and larger, with all XCFs converged to zero from 4QLs. This is consistent with the experimental report of Ref.~\cite{Liu-PRB-2012}. Comparing the trends that can be observed in Fig.~\ref{Figure5}, in which bandgap as a function of the number of QLs for both experimental and vdW-treated XCFs data is illustrated, we project that six QLs for Bi$_2$Se$_3$ and four QLs for Bi$_2$Te$_3$ are minimum models that capture a stable TSS.

Finally, in Fig. \ref{figure6}, we compare the band structures produced from models optimized with PBE and PBE+vdW for Bi$_2$Se$_3$ and Bi$_2$Te$_3$. The color coding of the band structure is based on the algorithm described in Fig.~\ref{figure7}, as in Ref.~\cite{Geilhufe-PRB-2015}, where $w$ is the average contribution of a type of atom for a given orbital to the band. For example, $w(\text{Bi}_p)$ is the weight of $p$-orbital of Bi atom in a specific band. It must be emphasized that for Bi$_2$Se$_3$ and Bi$_2$Te$_3$ the $p$ orbitals constitute the valence band ($v$) and the conduction band ($c$) states near the Fermi level. As can be seen, the vdW-corrected models produce the gapless $\Gamma$-point states and linear dispersion expected at 6QLs for both systems. In contrast, the PBE-optimized structures do not. Given that both models are derived from the same charge density predictions, this figure demonstrates the importance of the vdW treatments in producing results that agree with experiment purely as a matter of atomic structure.


\section{Conclusions}

Using prototypical TIs Bi$_2$Se$_3$ and Bi$_2$Te$_3$, we have here systematically assessed the role of several XCFs and relaxation effects, particularly as they relate to inter- and intra-QL separations, for prediction of TSS. We show here that there is an inextricable relationship between the number of quintuple layers, effects of structural optimization, and the electronic structure when different versions of XCFs are employed in first principles calculations carried out via DFT. A detailed analysis of the inter- and intra-layer relaxations reveal that GGA overestimates inter- and intra-QL separation, contradicting experimental findings~\cite{Roy-PRB-2014}.  On the other hand, the relatively simple LDA functional produces an opposite trend in relaxation effects, also in contradiction of experimental results. When more complicated XCFs such as metaGGA are employed, the relaxed structural parameters are found to be between the values obtained from experimental work and GGA. With vdW treatments applied, GGA and SCAN predict bulk-like crystal structures. In addition to structural parameters, we also investigate the thickness dependence of the bandgap for Bi$_2$Se$_3$ and Bi$_2$Te$_3$ for all XCFs. We ultimately determine that GGA+vdW offers the most reliable results, for both accuracy when compared to experimental studies and computability. We also determine that the GGA+vdW functional produces a structure that does not deviate much from the bulk, and generates a thickness-bandgap dependence that agrees well with experimental results. We thus conclude that structural optimization may not be necessary to model TI physics in binary layered systems of space group $R\bar{3}m$ (No. 166) within a tolerable accuracy.  It must be noted that modeling TI systems of this type with complex non-stoichiometric chemistry and doping effects would require structural optimization, where the GGA+vdW is most appropriate. Comparing the trends in bandgap as a function of number of quintuple layers for vdW-treated XCFs and experimental results, we conclude that the six QL model for Bi$_2$Se$_3$ and the four QL model for Bi$_2$Te$_3$ represent the minimum necessary thicknesses capable of capturing a stable TSS.

\section*{Acknowledgements}
The authors would like to acknowledge useful discussions with Matthias Geilhufe and Gayanath Fernando. AVB acknowledges the support of VILLUM FONDEN via the Center of Excellence for Dirac Materials (Grant No.11744) and the European Research Council under the European Unions Seventh Framework Program (FP/2207-2013)/ERC Grant Agreement No. DM-321031 ERC. 

The computational resources employed in this study were provided by High Performance Computing facility of University of Connecticut. 

\setcounter{table}{0}
\renewcommand{\thetable}{A\arabic{table}}

\newpage
\clearpage
\onecolumngrid
\appendix
\section{Appendix A}
\begin{table*}[!h]
\caption{\label{tab_em_BiSe} Table showing $\left|\frac{\partial E}{\partial k}\right|$ in eV/{\AA} and effective mass $m^*$ in units of rest mass of electron (refer Eqs.~(\ref{TaylorExpansionEnergy}) and (\ref{EffectiveMass})) of highest occupied band ($v$) and lowest unoccupied band ($c$) along the direction $\Gamma \rightarrow $M and $\Gamma \rightarrow $K of two-dimensional Brillouin zone for slab models of Bi$_2$Se$_3$ using various XCF treatment. The parity is estimated analyzing the $c$ and $v$ band occupancies following the algorithm discussed in Fig.~\ref{figure7}.} 
\begin{ruledtabular}
{\renewcommand{\arraystretch}{1.1}
\begin{tabular}{lcccccccccc}
\multirow{10}{*}{LDA} &
              Model &
              $P$ & 
              $ \left\vert\frac{\partial E}{\partial k}\right\vert_{\text{c}, \Gamma  \rightarrow M}$ & 
              $ \left\vert\frac{\partial E}{\partial k}\right\vert_{\text{c}, \Gamma  \rightarrow K}$ & 
              $ \left\vert\frac{\partial E}{\partial k}\right\vert_{\text{v}, \Gamma  \rightarrow M}$ & 
              $ \left\vert\frac{\partial E}{\partial k}\right\vert_{\text{v}, \Gamma \rightarrow K}$ & 
              $\vert m^{*}\vert _{\text{c}, \Gamma \rightarrow M}$ & 
              $\vert m^{*}\vert_{\text{c}, \Gamma \rightarrow K}$ & 
              $\vert m^{*}\vert_{\text{v}, \Gamma \rightarrow M}$ & 
              $\vert m^{*}\vert_{\text{v}, \Gamma \rightarrow K}$  \\[5pt]
           \hline      
  & 1QL & $+1$ & 0.99 & 1.12 & 1.12 & 1.27 & 0.15 & 0.16 & 0.15 & 0.17  \\
  & 2QL & $+1$ & 1.39 & 1.57 & 0.41 & 0.48 & 0.13 & 0.14 & 0.39 & 0.52  \\
  & 3QL & $-1$ & 1.86 & 2.00 & 0.06 & 0.03 & 0.15 & 0.16 & 0.29 & 0.42  \\
  & 4QL & $-1$ & 1.88 & 2.02 & 0.09 & 0.00 & 0.15 & 0.16 & 0.28 & 0.41  \\
  & 5QL & $-1$ & 1.95 & 2.09 & 0.16 & 0.07 & 0.16 & 0.16 & 0.26 & 0.37  \\
  & 6QL & $-1$ & 1.96 & 2.10 & 0.18 & 0.08 & 0.16 & 0.17 & 0.26 & 0.37  \\
  & 7QL & $-1$ & 1.97 & 2.10 & 0.19 & 0.09 & 0.16 & 0.17 & 0.26 & 0.37  \\
  & 8QL & $-1$ & 1.95 & 2.09 & 0.18 & 0.08 & 0.16 & 0.17 & 0.26 & 0.38  \\ 
  \hline      
 \multirow{8}{*}{PBE} 
  & 1QL & $+1$ & 0.87 & 0.99 & 0.95 & 1.08 & 0.16 & 0.17 & 0.17 & 0.19  \\
  & 2QL & $+1$ & 0.98 & 1.12 & 0.24 & 0.27 & 0.15 & 0.16 & 1.04 & 1.52  \\
  & 3QL & $+1$ & 1.02 & 1.16 & 0.13 & 0.15 & 0.14 & 0.15 & 1.78 & 2.62  \\
  & 4QL & $+1$ & 1.04 & 1.18 & 0.09 & 0.11 & 0.14 & 0.15 & 2.31 & 3.43  \\
  & 5QL & $+1$ & 1.10 & 1.26 & 0.04 & 0.04 & 0.14 & 0.15 & 2.68 & 2.23  \\
  & 6QL & $+1$ & 1.08 & 1.23 & 0.03 & 0.03 & 0.14 & 0.15 & 8.51 & 33.0  \\
  & 7QL & $+1$ & 1.27 & 1.44 & 0.29 & 0.32 & 0.13 & 0.14 & 0.68 & 0.71  \\
  & 8QL & $+1$ & 1.26 & 1.43 & 0.26 & 0.29 & 0.13 & 0.14 & 0.74 & 0.77  \\ 
  \hline      
 \multirow{8}{*}{SCAN} 
  & 1QL & $+1$ & 0.97 & 1.10 & 0.96 & 1.09 & 0.15 & 0.16 & 0.16 & 0.17  \\
  & 2QL & $+1$ & 1.13 & 1.29 & 0.36 & 0.40 & 0.13 & 0.14 & 0.84 & 1.47  \\
  & 3QL & $+1$ & 1.25 & 1.42 & 0.11 & 0.12 & 0.13 & 0.14 & 4.04 & 1.64  \\
  & 4QL & $+1$ & 1.50 & 1.68 & 0.11 & 0.12 & 0.13 & 0.14 & 1.49 & 0.91  \\
  & 5QL & $+1$ & 1.58 & 1.76 & 0.25 & 0.27 & 0.13 & 0.14 & 1.18 & 0.80  \\
  & 6QL & $+1$ & 1.67 & 1.84 & 0.41 & 0.43 & 0.13 & 0.15 & 1.07 & 0.81  \\
  & 7QL & $+1$ & 1.82 & 1.98 & 0.55 & 0.56 & 0.14 & 0.15 & 1.87 & 1.01  \\
  & 8QL & $+1$ & 1.86 & 2.02 & 0.59 & 0.59 & 0.15 & 0.16 & 2.25 & 1.05  \\   
    \hline      
 \multirow{8}{*}{PBE+vdW} 
  & 1QL & $+1$ & 0.87 & 0.99 & 0.98 & 1.11 & 0.16 & 0.17 & 0.17 & 0.19  \\
  & 2QL & $+1$ & 1.11 & 1.26 & 0.26 & 0.28 & 0.13 & 0.14 & 1.79 & 11.9  \\
  & 3QL & $+1$ & 1.45 & 1.62 & 0.14 & 0.14 & 0.13 & 0.14 & 1.87 & 1.02  \\
  & 4QL & $+1$ & 1.75 & 1.91 & 0.44 & 0.44 & 0.15 & 0.16 & 49.3 & 1.37  \\
  & 5QL & $+1$ & 1.88 & 2.03 & 0.62 & 0.61 & 0.15 & 0.17 & 11.6 & 1.44  \\
  & 6QL & $+1$ & 2.11 & 2.23 & 0.82 & 0.79 & 0.18 & 0.19 & 0.92 & 22.2  \\
  & 7QL & $+1$ & 2.08 & 2.21 & 0.79 & 0.76 & 0.18 & 0.19 & 0.96 & 21.3  \\
  & 8QL & $+1$ & 2.18 & 2.29 & 0.90 & 0.86 & 0.19 & 0.20 & 0.76 & 6.11  \\ 
  \hline      
 \multirow{8}{*}{SCAN+vdW} 
  & 1QL & $+1$ & 0.97 & 1.10 & 0.99 & 1.12 & 0.15 & 0.16 & 0.16 & 0.17  \\
  & 2QL & $+1$ & 1.25 & 1.42 & 0.42 & 0.47 & 0.12 & 0.13 & 0.62 & 1.04  \\
  & 3QL & $+1$ & 1.67 & 1.84 & 0.02 & 0.02 & 0.13 & 0.14 & 0.76 & 1.91  \\
  & 4QL & $-1$ & 1.90 & 2.05 & 0.26 & 0.21 & 0.14 & 0.16 & 0.53 & 1.35  \\
  & 5QL & $-1$ & 2.01 & 2.15 & 0.37 & 0.30 & 0.16 & 0.17 & 0.41 & 0.96  \\
  & 6QL & $-1$ & 2.07 & 2.20 & 0.44 & 0.37 & 0.16 & 0.17 & 0.40 & 0.94  \\
  & 7QL & $-1$ & 2.08 & 2.21 & 0.45 & 0.38 & 0.17 & 0.18 & 0.38 & 0.85  \\
  & 8QL & $-1$ & 2.10 & 2.23 & 0.48 & 0.40 & 0.17 & 0.18 & 0.39 & 0.85  \\             
\end{tabular}
}
\end{ruledtabular}
\end{table*}

\clearpage
\section{Appendix B}
\begin{table*}[!h]
\caption{\label{tab_em_BiTe} Table showing $\left|\frac{\partial E}{\partial k}\right|$ in eV/{\AA} and effective mass $m^*$ in units of rest mass of electron (refer Eqs.~(\ref{TaylorExpansionEnergy}) and (\ref{EffectiveMass})) of highest occupied band ($v$) and lowest unoccupied band ($c$) along the direction $\Gamma \rightarrow $M and $\Gamma \rightarrow $K of two-dimensional Brillouin zone for slab models of Bi$_2$Te$_3$ using various XCF treatment. The parity is estimated analyzing the $c$ and $v$ band occupancies following the algorithm discussed in Fig.~\ref{figure7}.}

\begin{ruledtabular}
{\renewcommand{\arraystretch}{1.1}
\begin{tabular}{lcccccccccc}
\multirow{10}{*}{LDA} &
              Model &
              $P$ & 
              $ \left\vert\frac{\partial E}{\partial k}\right\vert_{\text{c}, \Gamma  \rightarrow M}$ & 
              $ \left\vert\frac{\partial E}{\partial k}\right\vert_{\text{c}, \Gamma  \rightarrow K}$ & 
              $ \left\vert\frac{\partial E}{\partial k}\right\vert_{\text{v}, \Gamma  \rightarrow M}$ & 
              $ \left\vert\frac{\partial E}{\partial k}\right\vert_{\text{v}, \Gamma \rightarrow K}$ & 
              $\vert m^{*}\vert _{\text{c}, \Gamma \rightarrow M}$ & 
              $\vert m^{*}\vert_{\text{c}, \Gamma \rightarrow K}$ & 
              $\vert m^{*}\vert_{\text{v}, \Gamma \rightarrow M}$ & 
              $\vert m^{*}\vert_{\text{v}, \Gamma \rightarrow K}$  \\[5pt]
           \hline      
  & 1QL & $+1$ & 1.31 & 1.47 & 1.09 & 1.19 & 0.14 & 0.16 & 0.30 & 0.12  \\
  & 2QL & $-1$ & 1.13 & 1.27 & 1.54 & 1.64 & 0.17 & 0.18 & 0.53 & 8.42  \\
  & 3QL & $-1$ & 1.36 & 1.53 & 0.72 & 0.79 & 0.16 & 0.17 & 0.37 & 1.23  \\
  & 4QL & $-1$ & 1.68 & 1.82 & 0.41 & 0.52 & 0.21 & 0.21 & 0.20 & 0.37  \\
  & 5QL & $+1$ & 1.59 & 1.75 & 0.52 & 0.62 & 0.20 & 0.20 & 0.20 & 0.33  \\
  & 6QL & $+1$ & 1.67 & 1.82 & 0.44 & 0.55 & 0.21 & 0.21 & 0.18 & 0.29  \\
  & 7QL & $+1$ & 1.69 & 1.84 & 0.42 & 0.52 & 0.22 & 0.22 & 0.18 & 0.28  \\
  & 8QL & $+1$ & 1.70 & 1.84 & 0.40 & 0.51 & 0.22 & 0.22 & 0.18 & 0.28  \\ 
  \hline      
 \multirow{8}{*}{PBE} 
  & 1QL & $+1$ & 1.25 & 1.40 & 1.07 & 1.18 & 0.15 & 0.17 & 1.31 & 0.34  \\
  & 2QL & $+1$ & 2.25 & 2.41 & 0.44 & 0.41 & 0.22 & 0.26 & 0.95 & 2.81  \\
  & 3QL & $+1$ & 2.40 & 2.54 & 0.63 & 0.64 & 0.28 & 0.31 & 0.26 & 0.87  \\
  & 4QL & $+1$ & 2.74 & 2.87 & 1.02 & 1.04 & 0.45 & 0.42 & 0.20 & 0.98  \\
  & 5QL & $+1$ & 2.55 & 2.68 & 0.68 & 0.67 & 0.45 & 0.37 & 0.18 & 0.45  \\
  & 6QL & $-1$ & 2.86 & 2.99 & 1.08 & 1.10 & 1.12 & 0.51 & 0.15 & 0.39  \\
  & 7QL & $+1$ & 2.69 & 2.82 & 0.83 & 0.79 & 0.95 & 0.43 & 0.17 & 0.39  \\
  & 8QL & $+1$ & 2.81 & 2.93 & 1.01 & 0.99 & 3.09 & 0.50 & 0.14 & 0.32  \\ 
  \hline      
 \multirow{8}{*}{SCAN} 
  & 1QL & $+1$ & 2.13 & 2.29 & 0.04 & 0.09 & 0.19 & 0.23 & 0.93 & 14.7  \\
  & 2QL & $-1$ & 1.13 & 1.29 & 0.36 & 0.40 & 0.13 & 0.14 & 0.84 & 1.47  \\
  & 3QL & $+1$ & 1.99 & 2.17 & 0.22 & 0.25 & 0.19 & 0.20 & 0.27 & 1.34  \\
  & 4QL & $+1$ & 2.32 & 2.47 & 0.17 & 0.12 & 0.28 & 0.27 & 0.19 & 0.43  \\
  & 5QL & $-1$ & 2.42 & 2.56 & 0.27 & 0.21 & 0.35 & 0.30 & 0.19 & 0.44  \\
  & 6QL & $+1$ & 2.46 & 2.60 & 0.30 & 0.23 & 0.38 & 0.31 & 0.19 & 0.43  \\
  & 7QL & $+1$ & 2.42 & 2.62 & 0.31 & 0.24 & 0.42 & 0.31 & 0.19 & 0.43  \\
  & 8QL & $+1$ & 2.49 & 2.63 & 0.33 & 0.25 & 0.47 & 0.33 & 0.18 & 0.39  \\   
    \hline      
 \multirow{8}{*}{PBE+vdW} 
  & 1QL & $+1$ & 1.21 & 1.36 & 1.05 & 1.15 & 0.15 & 0.17 & 1.24 & 0.20  \\
  & 2QL & $+1$ & 1.36 & 1.53 & 0.79 & 0.83 & 0.16 & 0.18 & 2.79 & 1.05  \\
  & 3QL & $+1$ & 2.05 & 2.18 & 0.08 & 0.01 & 0.27 & 0.26 & 0.22 & 0.53  \\
  & 4QL & $+1$ & 1.83 & 1.99 & 0.19 & 0.25 & 0.22 & 0.22 & 0.22 & 0.50  \\
  & 5QL & $+1$ & 2.03 & 2.17 & 0.05 & 0.03 & 0.28 & 0.26 & 0.19 & 0.35  \\
  & 6QL & $+1$ & 2.09 & 2.22 & 0.10 & 0.01 & 0.30 & 0.27 & 0.19 & 0.35  \\
  & 7QL & $+1$ & 2.11 & 2.24 & 0.12 & 0.03 & 0.31 & 0.28 & 0.18 & 0.35  \\
  & 8QL & $+1$ & 2.05 & 2.19 & 0.05 & 0.03 & 0.29 & 0.26 & 0.20 & 0.38  \\ 
  \hline      
 \multirow{8}{*}{SCAN+vdW} 
  & 1QL & $+1$ & 1.30 & 1.45 & 1.10 & 1.20 & 0.14 & 0.16 & 0.60 & 0.16  \\
  & 2QL & $-1$ & 1.30 & 1.47 & 1.16 & 1.23 & 0.15 & 0.17 & 0.91 & 1.97  \\
  & 3QL & $+1$ & 1.89 & 2.05 & 0.26 & 0.33 & 0.20 & 0.20 & 0.25 & 0.65  \\
  & 4QL & $+1$ & 1.82 & 1.99 & 0.39 & 0.47 & 0.19 & 0.20 & 0.21 & 0.45  \\
  & 5QL & $+1$ & 2.01 & 2.16 & 0.18 & 0.27 & 0.23 & 0.23 & 0.18 & 0.32  \\
  & 6QL & $+1$ & 2.09 & 2.23 & 0.09 & 0.20 & 0.25 & 0.24 & 0.17 & 0.29  \\
  & 7QL & $+1$ & 2.07 & 2.21 & 0.11 & 0.22 & 0.25 & 0.24 & 0.17 & 0.29  \\
  & 8QL & $+1$ & 2.14 & 2.28 & 0.05 & 0.16 & 0.27 & 0.25 & 0.17 & 0.32  \\             
\end{tabular}
}
\end{ruledtabular}
\end{table*}

\newpage
\clearpage
\twocolumngrid
\bibliographystyle{apsrev4-1}
\bibliography{references}

\begin{thebibliography}{64}%
\makeatletter
\providecommand \@ifxundefined [1]{%
 \@ifx{#1\undefined}
}%
\providecommand \@ifnum [1]{%
 \ifnum #1\expandafter \@firstoftwo
 \else \expandafter \@secondoftwo
 \fi
}%
\providecommand \@ifx [1]{%
 \ifx #1\expandafter \@firstoftwo
 \else \expandafter \@secondoftwo
 \fi
}%
\providecommand \natexlab [1]{#1}%
\providecommand \enquote  [1]{``#1''}%
\providecommand \bibnamefont  [1]{#1}%
\providecommand \bibfnamefont [1]{#1}%
\providecommand \citenamefont [1]{#1}%
\providecommand \href@noop [0]{\@secondoftwo}%
\providecommand \href [0]{\begingroup \@sanitize@url \@href}%
\providecommand \@href[1]{\@@startlink{#1}\@@href}%
\providecommand \@@href[1]{\endgroup#1\@@endlink}%
\providecommand \@sanitize@url [0]{\catcode `\\12\catcode `\$12\catcode
  `\&12\catcode `\#12\catcode `\^12\catcode `\_12\catcode `\%12\relax}%
\providecommand \@@startlink[1]{}%
\providecommand \@@endlink[0]{}%
\providecommand \url  [0]{\begingroup\@sanitize@url \@url }%
\providecommand \@url [1]{\endgroup\@href {#1}{\urlprefix }}%
\providecommand \urlprefix  [0]{URL }%
\providecommand \Eprint [0]{\href }%
\providecommand \doibase [0]{http://dx.doi.org/}%
\providecommand \selectlanguage [0]{\@gobble}%
\providecommand \bibinfo  [0]{\@secondoftwo}%
\providecommand \bibfield  [0]{\@secondoftwo}%
\providecommand \translation [1]{[#1]}%
\providecommand \BibitemOpen [0]{}%
\providecommand \bibitemStop [0]{}%
\providecommand \bibitemNoStop [0]{.\EOS\space}%
\providecommand \EOS [0]{\spacefactor3000\relax}%
\providecommand \BibitemShut  [1]{\csname bibitem#1\endcsname}%
\let\auto@bib@innerbib\@empty
\bibitem [{\citenamefont {Zhang}\ \emph {et~al.}(2009)\citenamefont {Zhang},
  \citenamefont {Liu}, \citenamefont {Qi}, \citenamefont {Dai}, \citenamefont
  {Fang},\ and\ \citenamefont {Zhang}}]{Zhang-NP-2009}%
  \BibitemOpen
  \bibfield  {author} {\bibinfo {author} {\bibfnamefont {H.}~\bibnamefont
  {Zhang}}, \bibinfo {author} {\bibfnamefont {C.-X.}\ \bibnamefont {Liu}},
  \bibinfo {author} {\bibfnamefont {X.-L.}\ \bibnamefont {Qi}}, \bibinfo
  {author} {\bibfnamefont {X.}~\bibnamefont {Dai}}, \bibinfo {author}
  {\bibfnamefont {Z.}~\bibnamefont {Fang}}, \ and\ \bibinfo {author}
  {\bibfnamefont {S.-C.}\ \bibnamefont {Zhang}},\ }\href {\doibase
  https://doi.org/10.1038/nphys1270} {\bibfield  {journal} {\bibinfo  {journal}
  {Nature Phys.}\ }\textbf {\bibinfo {volume} {5}},\ \bibinfo {pages} {438}
  (\bibinfo {year} {2009})}\BibitemShut {NoStop}%
\bibitem [{\citenamefont {Xia}\ \emph {et~al.}(2009)\citenamefont {Xia},
  \citenamefont {Qian}, \citenamefont {Hsieh}, \citenamefont {Wray},
  \citenamefont {Pal}, \citenamefont {Lin}, \citenamefont {Bansil},
  \citenamefont {Grauer}, \citenamefont {Hor}, \citenamefont {Cava},\ and\
  \citenamefont {Hasan}}]{Xia-NP-2009}%
  \BibitemOpen
  \bibfield  {author} {\bibinfo {author} {\bibfnamefont {Y.}~\bibnamefont
  {Xia}}, \bibinfo {author} {\bibfnamefont {D.}~\bibnamefont {Qian}}, \bibinfo
  {author} {\bibfnamefont {D.}~\bibnamefont {Hsieh}}, \bibinfo {author}
  {\bibfnamefont {L.}~\bibnamefont {Wray}}, \bibinfo {author} {\bibfnamefont
  {A.}~\bibnamefont {Pal}}, \bibinfo {author} {\bibfnamefont {H.}~\bibnamefont
  {Lin}}, \bibinfo {author} {\bibfnamefont {A.}~\bibnamefont {Bansil}},
  \bibinfo {author} {\bibfnamefont {D.}~\bibnamefont {Grauer}}, \bibinfo
  {author} {\bibfnamefont {Y.~S.}\ \bibnamefont {Hor}}, \bibinfo {author}
  {\bibfnamefont {R.~J.}\ \bibnamefont {Cava}}, \ and\ \bibinfo {author}
  {\bibfnamefont {M.~Z.}\ \bibnamefont {Hasan}},\ }\href {\doibase
  https://doi.org/10.1038/nphys1274} {\bibfield  {journal} {\bibinfo  {journal}
  {Nature Phys.}\ }\textbf {\bibinfo {volume} {5}},\ \bibinfo {pages} {398}
  (\bibinfo {year} {2009})}\BibitemShut {NoStop}%
\bibitem [{\citenamefont {Hsieh}\ \emph
  {et~al.}(2009{\natexlab{a}})\citenamefont {Hsieh}, \citenamefont {Xia},
  \citenamefont {Qian}, \citenamefont {Wray}, \citenamefont {Dil},
  \citenamefont {Meier}, \citenamefont {Osterwalder}, \citenamefont {Patthey},
  \citenamefont {Checkelsky}, \citenamefont {Ong}, \citenamefont {Fedorov},
  \citenamefont {Lin}, \citenamefont {Bansil}, \citenamefont {Grauer},
  \citenamefont {Hor}, \citenamefont {Cava},\ and\ \citenamefont
  {Hasan}}]{Hsieh-Nature-2009}%
  \BibitemOpen
  \bibfield  {author} {\bibinfo {author} {\bibfnamefont {D.}~\bibnamefont
  {Hsieh}}, \bibinfo {author} {\bibfnamefont {Y.}~\bibnamefont {Xia}}, \bibinfo
  {author} {\bibfnamefont {D.}~\bibnamefont {Qian}}, \bibinfo {author}
  {\bibfnamefont {L.}~\bibnamefont {Wray}}, \bibinfo {author} {\bibfnamefont
  {J.~H.}\ \bibnamefont {Dil}}, \bibinfo {author} {\bibfnamefont
  {F.}~\bibnamefont {Meier}}, \bibinfo {author} {\bibfnamefont
  {J.}~\bibnamefont {Osterwalder}}, \bibinfo {author} {\bibfnamefont
  {L.}~\bibnamefont {Patthey}}, \bibinfo {author} {\bibfnamefont {J.~G.}\
  \bibnamefont {Checkelsky}}, \bibinfo {author} {\bibfnamefont {N.~P.}\
  \bibnamefont {Ong}}, \bibinfo {author} {\bibfnamefont {A.~V.}\ \bibnamefont
  {Fedorov}}, \bibinfo {author} {\bibfnamefont {H.}~\bibnamefont {Lin}},
  \bibinfo {author} {\bibfnamefont {A.}~\bibnamefont {Bansil}}, \bibinfo
  {author} {\bibfnamefont {D.}~\bibnamefont {Grauer}}, \bibinfo {author}
  {\bibfnamefont {Y.~S.}\ \bibnamefont {Hor}}, \bibinfo {author} {\bibfnamefont
  {R.~J.}\ \bibnamefont {Cava}}, \ and\ \bibinfo {author} {\bibfnamefont
  {M.~Z.}\ \bibnamefont {Hasan}},\ }\href {\doibase
  https://doi.org/10.1038/nature08234} {\bibfield  {journal} {\bibinfo
  {journal} {Nature}\ }\textbf {\bibinfo {volume} {460}},\ \bibinfo {pages}
  {1101} (\bibinfo {year} {2009}{\natexlab{a}})}\BibitemShut {NoStop}%
\bibitem [{\citenamefont {Hasan}\ and\ \citenamefont
  {Kane}(2010)}]{Hasan-RMP-2010}%
  \BibitemOpen
  \bibfield  {author} {\bibinfo {author} {\bibfnamefont {M.~Z.}\ \bibnamefont
  {Hasan}}\ and\ \bibinfo {author} {\bibfnamefont {C.~L.}\ \bibnamefont
  {Kane}},\ }\href {\doibase 10.1103/RevModPhys.82.3045} {\bibfield  {journal}
  {\bibinfo  {journal} {Rev. Mod. Phys.}\ }\textbf {\bibinfo {volume} {82}},\
  \bibinfo {pages} {3045} (\bibinfo {year} {2010})}\BibitemShut {NoStop}%
\bibitem [{\citenamefont {Moore}(2010)}]{Moore-Nature-2010}%
  \BibitemOpen
  \bibfield  {author} {\bibinfo {author} {\bibfnamefont {J.~E.}\ \bibnamefont
  {Moore}},\ }\href {\doibase https://doi.org/10.1038/nature08916} {\bibfield
  {journal} {\bibinfo  {journal} {Nature}\ }\textbf {\bibinfo {volume} {464}},\
  \bibinfo {pages} {194} (\bibinfo {year} {2010})}\BibitemShut {NoStop}%
\bibitem [{\citenamefont {Zhang}\ \emph {et~al.}(2010)\citenamefont {Zhang},
  \citenamefont {He}, \citenamefont {Chang}, \citenamefont {Song},
  \citenamefont {Wang}, \citenamefont {Chen}, \citenamefont {Jia},
  \citenamefont {Fang}, \citenamefont {Dai}, \citenamefont {Shan},
  \citenamefont {Shen}, \citenamefont {Niu}, \citenamefont {Qi}, \citenamefont
  {Zhang}, \citenamefont {Ma},\ and\ \citenamefont {Xue}}]{Zhang-NP-2010}%
  \BibitemOpen
  \bibfield  {author} {\bibinfo {author} {\bibfnamefont {Y.}~\bibnamefont
  {Zhang}}, \bibinfo {author} {\bibfnamefont {K.}~\bibnamefont {He}}, \bibinfo
  {author} {\bibfnamefont {C.-Z.}\ \bibnamefont {Chang}}, \bibinfo {author}
  {\bibfnamefont {C.-L.}\ \bibnamefont {Song}}, \bibinfo {author}
  {\bibfnamefont {L.-L.}\ \bibnamefont {Wang}}, \bibinfo {author}
  {\bibfnamefont {X.}~\bibnamefont {Chen}}, \bibinfo {author} {\bibfnamefont
  {J.-F.}\ \bibnamefont {Jia}}, \bibinfo {author} {\bibfnamefont
  {Z.}~\bibnamefont {Fang}}, \bibinfo {author} {\bibfnamefont {X.}~\bibnamefont
  {Dai}}, \bibinfo {author} {\bibfnamefont {W.-Y.}\ \bibnamefont {Shan}},
  \bibinfo {author} {\bibfnamefont {S.-Q.}\ \bibnamefont {Shen}}, \bibinfo
  {author} {\bibfnamefont {Q.}~\bibnamefont {Niu}}, \bibinfo {author}
  {\bibfnamefont {X.-L.}\ \bibnamefont {Qi}}, \bibinfo {author} {\bibfnamefont
  {S.-C.}\ \bibnamefont {Zhang}}, \bibinfo {author} {\bibfnamefont {X.-C.}\
  \bibnamefont {Ma}}, \ and\ \bibinfo {author} {\bibfnamefont {Q.-K.}\
  \bibnamefont {Xue}},\ }\href {\doibase https://doi.org/10.1038/nphys1689}
  {\bibfield  {journal} {\bibinfo  {journal} {Nature Phys.}\ }\textbf {\bibinfo
  {volume} {6}},\ \bibinfo {pages} {584} (\bibinfo {year} {2010})}\BibitemShut
  {NoStop}%
\bibitem [{\citenamefont {Taskin}\ \emph {et~al.}(2012)\citenamefont {Taskin},
  \citenamefont {Sasaki}, \citenamefont {Segawa},\ and\ \citenamefont
  {Ando}}]{Taskin-AM-2012}%
  \BibitemOpen
  \bibfield  {author} {\bibinfo {author} {\bibfnamefont {A.~A.}\ \bibnamefont
  {Taskin}}, \bibinfo {author} {\bibfnamefont {S.}~\bibnamefont {Sasaki}},
  \bibinfo {author} {\bibfnamefont {K.}~\bibnamefont {Segawa}}, \ and\ \bibinfo
  {author} {\bibfnamefont {Y.}~\bibnamefont {Ando}},\ }\href {\doibase
  10.1002/adma.201201827} {\bibfield  {journal} {\bibinfo  {journal} {Advanced
  Materials}\ }\textbf {\bibinfo {volume} {24}},\ \bibinfo {pages} {5581}
  (\bibinfo {year} {2012})}\BibitemShut {NoStop}%
\bibitem [{\citenamefont {Tsipas}\ \emph {et~al.}(2014)\citenamefont {Tsipas},
  \citenamefont {Xenogiannopoulou}, \citenamefont {Kassavetis}, \citenamefont
  {Tsoutsou}, \citenamefont {Golias}, \citenamefont {Bazioti}, \citenamefont
  {Dimitrakopulos}, \citenamefont {Komninou}, \citenamefont {Liang},
  \citenamefont {Caymax},\ and\ \citenamefont {Dimoulas}}]{Tsipas-ACSN-2014}%
  \BibitemOpen
  \bibfield  {author} {\bibinfo {author} {\bibfnamefont {P.}~\bibnamefont
  {Tsipas}}, \bibinfo {author} {\bibfnamefont {E.}~\bibnamefont
  {Xenogiannopoulou}}, \bibinfo {author} {\bibfnamefont {S.}~\bibnamefont
  {Kassavetis}}, \bibinfo {author} {\bibfnamefont {D.}~\bibnamefont
  {Tsoutsou}}, \bibinfo {author} {\bibfnamefont {E.}~\bibnamefont {Golias}},
  \bibinfo {author} {\bibfnamefont {C.}~\bibnamefont {Bazioti}}, \bibinfo
  {author} {\bibfnamefont {G.~P.}\ \bibnamefont {Dimitrakopulos}}, \bibinfo
  {author} {\bibfnamefont {P.}~\bibnamefont {Komninou}}, \bibinfo {author}
  {\bibfnamefont {H.}~\bibnamefont {Liang}}, \bibinfo {author} {\bibfnamefont
  {M.}~\bibnamefont {Caymax}}, \ and\ \bibinfo {author} {\bibfnamefont
  {A.}~\bibnamefont {Dimoulas}},\ }\href {\doibase 10.1021/nn502397x}
  {\bibfield  {journal} {\bibinfo  {journal} {ACS Nano}\ }\textbf {\bibinfo
  {volume} {8}},\ \bibinfo {pages} {6614} (\bibinfo {year} {2014})}\BibitemShut
  {NoStop}%
\bibitem [{\citenamefont {Das~Sarma}\ \emph {et~al.}(2006)\citenamefont
  {Das~Sarma}, \citenamefont {Freedman},\ and\ \citenamefont
  {Nayak}}]{Das-Sarma-PhyToday-2006}%
  \BibitemOpen
  \bibfield  {author} {\bibinfo {author} {\bibfnamefont {S.}~\bibnamefont
  {Das~Sarma}}, \bibinfo {author} {\bibfnamefont {M.}~\bibnamefont {Freedman}},
  \ and\ \bibinfo {author} {\bibfnamefont {C.}~\bibnamefont {Nayak}},\ }\href
  {\doibase 10.1063/1.2337825} {\bibfield  {journal} {\bibinfo  {journal}
  {Physics Today}\ }\textbf {\bibinfo {volume} {59}},\ \bibinfo {pages} {32}
  (\bibinfo {year} {2006})}\BibitemShut {NoStop}%
\bibitem [{\citenamefont {Wang}\ \emph {et~al.}(2016)\citenamefont {Wang},
  \citenamefont {Lang},\ and\ \citenamefont {Kou}}]{Wang-2016}%
  \BibitemOpen
  \bibfield  {author} {\bibinfo {author} {\bibfnamefont {K.~L.}\ \bibnamefont
  {Wang}}, \bibinfo {author} {\bibfnamefont {M.}~\bibnamefont {Lang}}, \ and\
  \bibinfo {author} {\bibfnamefont {X.}~\bibnamefont {Kou}},\ }\enquote
  {\bibinfo {title} {Spintronics of topological insulators},}\ in\ \href
  {\doibase 10.1007/978-94-007-6892-5_56} {\emph {\bibinfo {booktitle}
  {Handbook of Spintronics}}},\ \bibinfo {editor} {edited by\ \bibinfo {editor}
  {\bibfnamefont {Y.}~\bibnamefont {Xu}}, \bibinfo {editor} {\bibfnamefont
  {D.~D.}\ \bibnamefont {Awschalom}}, \ and\ \bibinfo {editor} {\bibfnamefont
  {J.}~\bibnamefont {Nitta}}}\ (\bibinfo  {publisher} {Springer Netherlands},\
  \bibinfo {address} {Dordrecht},\ \bibinfo {year} {2016})\ pp.\ \bibinfo
  {pages} {431--462}\BibitemShut {NoStop}%
\bibitem [{\citenamefont {Fan}\ and\ \citenamefont
  {Wang}(2016)}]{Fan-Spin-2016}%
  \BibitemOpen
  \bibfield  {author} {\bibinfo {author} {\bibfnamefont {Y.}~\bibnamefont
  {Fan}}\ and\ \bibinfo {author} {\bibfnamefont {K.~L.}\ \bibnamefont {Wang}},\
  }\href {\doibase 10.1142/S2010324716400014} {\bibfield  {journal} {\bibinfo
  {journal} {SPIN}\ }\textbf {\bibinfo {volume} {06}},\ \bibinfo {pages}
  {1640001} (\bibinfo {year} {2016})}\BibitemShut {NoStop}%
\bibitem [{\citenamefont {Zhang}\ \emph
  {et~al.}(2013{\natexlab{a}})\citenamefont {Zhang}, \citenamefont {Kane},\
  and\ \citenamefont {Mele}}]{Zhang-PRL-2013-A}%
  \BibitemOpen
  \bibfield  {author} {\bibinfo {author} {\bibfnamefont {F.}~\bibnamefont
  {Zhang}}, \bibinfo {author} {\bibfnamefont {C.~L.}\ \bibnamefont {Kane}}, \
  and\ \bibinfo {author} {\bibfnamefont {E.~J.}\ \bibnamefont {Mele}},\ }\href
  {\doibase 10.1103/PhysRevLett.110.046404} {\bibfield  {journal} {\bibinfo
  {journal} {Phys. Rev. Lett.}\ }\textbf {\bibinfo {volume} {110}},\ \bibinfo
  {pages} {046404} (\bibinfo {year} {2013}{\natexlab{a}})}\BibitemShut
  {NoStop}%
\bibitem [{\citenamefont {Zhang}\ \emph
  {et~al.}(2013{\natexlab{b}})\citenamefont {Zhang}, \citenamefont {Kane},\
  and\ \citenamefont {Mele}}]{Zhang-PRL-2013-B}%
  \BibitemOpen
  \bibfield  {author} {\bibinfo {author} {\bibfnamefont {F.}~\bibnamefont
  {Zhang}}, \bibinfo {author} {\bibfnamefont {C.~L.}\ \bibnamefont {Kane}}, \
  and\ \bibinfo {author} {\bibfnamefont {E.~J.}\ \bibnamefont {Mele}},\ }\href
  {\doibase 10.1103/PhysRevLett.111.056403} {\bibfield  {journal} {\bibinfo
  {journal} {Phys. Rev. Lett.}\ }\textbf {\bibinfo {volume} {111}},\ \bibinfo
  {pages} {056403} (\bibinfo {year} {2013}{\natexlab{b}})}\BibitemShut
  {NoStop}%
\bibitem [{\citenamefont {Kane}\ and\ \citenamefont
  {Zhang}(2015)}]{Kane-PhysScripta-2015}%
  \BibitemOpen
  \bibfield  {author} {\bibinfo {author} {\bibfnamefont {C.~L.}\ \bibnamefont
  {Kane}}\ and\ \bibinfo {author} {\bibfnamefont {F.}~\bibnamefont {Zhang}},\
  }\href {\doibase 10.1088/0031-8949/2015/t164/014011} {\bibfield  {journal}
  {\bibinfo  {journal} {Physica Scripta}\ }\textbf {\bibinfo {volume} {T164}},\
  \bibinfo {pages} {014011} (\bibinfo {year} {2015})}\BibitemShut {NoStop}%
\bibitem [{\citenamefont {Metlitski}\ \emph {et~al.}(2015)\citenamefont
  {Metlitski}, \citenamefont {Kane},\ and\ \citenamefont
  {Fisher}}]{Metlitski-PRB-2015}%
  \BibitemOpen
  \bibfield  {author} {\bibinfo {author} {\bibfnamefont {M.~A.}\ \bibnamefont
  {Metlitski}}, \bibinfo {author} {\bibfnamefont {C.~L.}\ \bibnamefont {Kane}},
  \ and\ \bibinfo {author} {\bibfnamefont {M.~P.~A.}\ \bibnamefont {Fisher}},\
  }\href {\doibase 10.1103/PhysRevB.92.125111} {\bibfield  {journal} {\bibinfo
  {journal} {Phys. Rev. B}\ }\textbf {\bibinfo {volume} {92}},\ \bibinfo
  {pages} {125111} (\bibinfo {year} {2015})}\BibitemShut {NoStop}%
\bibitem [{\citenamefont {Teo}\ \emph {et~al.}(2008)\citenamefont {Teo},
  \citenamefont {Fu},\ and\ \citenamefont {Kane}}]{Teo-PRB-2008}%
  \BibitemOpen
  \bibfield  {author} {\bibinfo {author} {\bibfnamefont {J.~C.~Y.}\
  \bibnamefont {Teo}}, \bibinfo {author} {\bibfnamefont {L.}~\bibnamefont
  {Fu}}, \ and\ \bibinfo {author} {\bibfnamefont {C.~L.}\ \bibnamefont
  {Kane}},\ }\href {\doibase 10.1103/PhysRevB.78.045426} {\bibfield  {journal}
  {\bibinfo  {journal} {Phys. Rev. B}\ }\textbf {\bibinfo {volume} {78}},\
  \bibinfo {pages} {045426} (\bibinfo {year} {2008})}\BibitemShut {NoStop}%
\bibitem [{\citenamefont {Hsieh}\ \emph
  {et~al.}(2009{\natexlab{b}})\citenamefont {Hsieh}, \citenamefont {Xia},
  \citenamefont {Wray}, \citenamefont {Qian}, \citenamefont {Pal},
  \citenamefont {Dil}, \citenamefont {Osterwalder}, \citenamefont {Meier},
  \citenamefont {Bihlmayer}, \citenamefont {Kane}, \citenamefont {Hor},
  \citenamefont {Cava},\ and\ \citenamefont {Hasan}}]{Hsieh-Science-2009}%
  \BibitemOpen
  \bibfield  {author} {\bibinfo {author} {\bibfnamefont {D.}~\bibnamefont
  {Hsieh}}, \bibinfo {author} {\bibfnamefont {Y.}~\bibnamefont {Xia}}, \bibinfo
  {author} {\bibfnamefont {L.}~\bibnamefont {Wray}}, \bibinfo {author}
  {\bibfnamefont {D.}~\bibnamefont {Qian}}, \bibinfo {author} {\bibfnamefont
  {A.}~\bibnamefont {Pal}}, \bibinfo {author} {\bibfnamefont {J.~H.}\
  \bibnamefont {Dil}}, \bibinfo {author} {\bibfnamefont {J.}~\bibnamefont
  {Osterwalder}}, \bibinfo {author} {\bibfnamefont {F.}~\bibnamefont {Meier}},
  \bibinfo {author} {\bibfnamefont {G.}~\bibnamefont {Bihlmayer}}, \bibinfo
  {author} {\bibfnamefont {C.~L.}\ \bibnamefont {Kane}}, \bibinfo {author}
  {\bibfnamefont {Y.~S.}\ \bibnamefont {Hor}}, \bibinfo {author} {\bibfnamefont
  {R.~J.}\ \bibnamefont {Cava}}, \ and\ \bibinfo {author} {\bibfnamefont
  {M.~Z.}\ \bibnamefont {Hasan}},\ }\href {\doibase 10.1126/science.1167733}
  {\bibfield  {journal} {\bibinfo  {journal} {Science}\ }\textbf {\bibinfo
  {volume} {323}},\ \bibinfo {pages} {919} (\bibinfo {year}
  {2009}{\natexlab{b}})}\BibitemShut {NoStop}%
\bibitem [{\citenamefont {Fu}\ and\ \citenamefont
  {Kane}(2008)}]{Liang-PRL-2008}%
  \BibitemOpen
  \bibfield  {author} {\bibinfo {author} {\bibfnamefont {L.}~\bibnamefont
  {Fu}}\ and\ \bibinfo {author} {\bibfnamefont {C.~L.}\ \bibnamefont {Kane}},\
  }\href {\doibase 10.1103/PhysRevLett.100.096407} {\bibfield  {journal}
  {\bibinfo  {journal} {Phys. Rev. Lett.}\ }\textbf {\bibinfo {volume} {100}},\
  \bibinfo {pages} {096407} (\bibinfo {year} {2008})}\BibitemShut {NoStop}%
\bibitem [{\citenamefont {Kane}\ and\ \citenamefont
  {Mele}(2005)}]{Kane-PRL-2005}%
  \BibitemOpen
  \bibfield  {author} {\bibinfo {author} {\bibfnamefont {C.~L.}\ \bibnamefont
  {Kane}}\ and\ \bibinfo {author} {\bibfnamefont {E.~J.}\ \bibnamefont
  {Mele}},\ }\href {\doibase 10.1103/PhysRevLett.95.146802} {\bibfield
  {journal} {\bibinfo  {journal} {Phys. Rev. Lett.}\ }\textbf {\bibinfo
  {volume} {95}},\ \bibinfo {pages} {146802} (\bibinfo {year}
  {2005})}\BibitemShut {NoStop}%
\bibitem [{\citenamefont {Wehling}\ \emph {et~al.}(2014)\citenamefont
  {Wehling}, \citenamefont {Black-Schaffer},\ and\ \citenamefont
  {Balatsky}}]{Wehling-AP-2014}%
  \BibitemOpen
  \bibfield  {author} {\bibinfo {author} {\bibfnamefont {T.}~\bibnamefont
  {Wehling}}, \bibinfo {author} {\bibfnamefont {A.}~\bibnamefont
  {Black-Schaffer}}, \ and\ \bibinfo {author} {\bibfnamefont {A.}~\bibnamefont
  {Balatsky}},\ }\href {\doibase 10.1080/00018732.2014.927109} {\bibfield
  {journal} {\bibinfo  {journal} {Advances in Physics}\ }\textbf {\bibinfo
  {volume} {63}},\ \bibinfo {pages} {1} (\bibinfo {year} {2014})}\BibitemShut
  {NoStop}%
\bibitem [{\citenamefont {Nakajima}(1963)}]{Nakajima-JPCS-1963}%
  \BibitemOpen
  \bibfield  {author} {\bibinfo {author} {\bibfnamefont {S.}~\bibnamefont
  {Nakajima}},\ }\href {\doibase https://doi.org/10.1016/0022-3697(63)90207-5}
  {\bibfield  {journal} {\bibinfo  {journal} {Journal of Physics and Chemistry
  of Solids}\ }\textbf {\bibinfo {volume} {24}},\ \bibinfo {pages} {479}
  (\bibinfo {year} {1963})}\BibitemShut {NoStop}%
\bibitem [{\citenamefont {Liu}\ \emph {et~al.}(2012)\citenamefont {Liu},
  \citenamefont {Bian}, \citenamefont {Miller}, \citenamefont {Bissen},\ and\
  \citenamefont {Chiang}}]{Liu-PRB-2012}%
  \BibitemOpen
  \bibfield  {author} {\bibinfo {author} {\bibfnamefont {Y.}~\bibnamefont
  {Liu}}, \bibinfo {author} {\bibfnamefont {G.}~\bibnamefont {Bian}}, \bibinfo
  {author} {\bibfnamefont {T.}~\bibnamefont {Miller}}, \bibinfo {author}
  {\bibfnamefont {M.}~\bibnamefont {Bissen}}, \ and\ \bibinfo {author}
  {\bibfnamefont {T.-C.}\ \bibnamefont {Chiang}},\ }\href {\doibase
  10.1103/PhysRevB.85.195442} {\bibfield  {journal} {\bibinfo  {journal} {Phys.
  Rev. B}\ }\textbf {\bibinfo {volume} {85}},\ \bibinfo {pages} {195442}
  (\bibinfo {year} {2012})}\BibitemShut {NoStop}%
\bibitem [{\citenamefont {Liu}\ \emph {et~al.}(2010)\citenamefont {Liu},
  \citenamefont {Zhang}, \citenamefont {Yan}, \citenamefont {Qi}, \citenamefont
  {Frauenheim}, \citenamefont {Dai}, \citenamefont {Fang},\ and\ \citenamefont
  {Zhang}}]{Chao-Xing-PRB-2010}%
  \BibitemOpen
  \bibfield  {author} {\bibinfo {author} {\bibfnamefont {C.-X.}\ \bibnamefont
  {Liu}}, \bibinfo {author} {\bibfnamefont {H.}~\bibnamefont {Zhang}}, \bibinfo
  {author} {\bibfnamefont {B.}~\bibnamefont {Yan}}, \bibinfo {author}
  {\bibfnamefont {X.-L.}\ \bibnamefont {Qi}}, \bibinfo {author} {\bibfnamefont
  {T.}~\bibnamefont {Frauenheim}}, \bibinfo {author} {\bibfnamefont
  {X.}~\bibnamefont {Dai}}, \bibinfo {author} {\bibfnamefont {Z.}~\bibnamefont
  {Fang}}, \ and\ \bibinfo {author} {\bibfnamefont {S.-C.}\ \bibnamefont
  {Zhang}},\ }\href {\doibase 10.1103/PhysRevB.81.041307} {\bibfield  {journal}
  {\bibinfo  {journal} {Phys. Rev. B}\ }\textbf {\bibinfo {volume} {81}},\
  \bibinfo {pages} {041307} (\bibinfo {year} {2010})}\BibitemShut {NoStop}%
\bibitem [{\citenamefont {Young}\ \emph {et~al.}(2011)\citenamefont {Young},
  \citenamefont {Chowdhury}, \citenamefont {Walter}, \citenamefont {Mele},
  \citenamefont {Kane},\ and\ \citenamefont {Rappe}}]{Young-PRB-2011}%
  \BibitemOpen
  \bibfield  {author} {\bibinfo {author} {\bibfnamefont {S.~M.}\ \bibnamefont
  {Young}}, \bibinfo {author} {\bibfnamefont {S.}~\bibnamefont {Chowdhury}},
  \bibinfo {author} {\bibfnamefont {E.~J.}\ \bibnamefont {Walter}}, \bibinfo
  {author} {\bibfnamefont {E.~J.}\ \bibnamefont {Mele}}, \bibinfo {author}
  {\bibfnamefont {C.~L.}\ \bibnamefont {Kane}}, \ and\ \bibinfo {author}
  {\bibfnamefont {A.~M.}\ \bibnamefont {Rappe}},\ }\href {\doibase
  10.1103/PhysRevB.84.085106} {\bibfield  {journal} {\bibinfo  {journal} {Phys.
  Rev. B}\ }\textbf {\bibinfo {volume} {84}},\ \bibinfo {pages} {085106}
  (\bibinfo {year} {2011})}\BibitemShut {NoStop}%
\bibitem [{\citenamefont {Liu}\ \emph {et~al.}(2011)\citenamefont {Liu},
  \citenamefont {Peng}, \citenamefont {Tang}, \citenamefont {Sun},
  \citenamefont {Zhang},\ and\ \citenamefont {Zhong}}]{Liu-PRB-2011}%
  \BibitemOpen
  \bibfield  {author} {\bibinfo {author} {\bibfnamefont {W.}~\bibnamefont
  {Liu}}, \bibinfo {author} {\bibfnamefont {X.}~\bibnamefont {Peng}}, \bibinfo
  {author} {\bibfnamefont {C.}~\bibnamefont {Tang}}, \bibinfo {author}
  {\bibfnamefont {L.}~\bibnamefont {Sun}}, \bibinfo {author} {\bibfnamefont
  {K.}~\bibnamefont {Zhang}}, \ and\ \bibinfo {author} {\bibfnamefont
  {J.}~\bibnamefont {Zhong}},\ }\href {\doibase 10.1103/PhysRevB.84.245105}
  {\bibfield  {journal} {\bibinfo  {journal} {Phys. Rev. B}\ }\textbf {\bibinfo
  {volume} {84}},\ \bibinfo {pages} {245105} (\bibinfo {year}
  {2011})}\BibitemShut {NoStop}%
\bibitem [{\citenamefont {Cohen}\ \emph {et~al.}(2008)\citenamefont {Cohen},
  \citenamefont {Mori-S{\'a}nchez},\ and\ \citenamefont
  {Yang}}]{Cohen-Science-2008}%
  \BibitemOpen
  \bibfield  {author} {\bibinfo {author} {\bibfnamefont {A.~J.}\ \bibnamefont
  {Cohen}}, \bibinfo {author} {\bibfnamefont {P.}~\bibnamefont
  {Mori-S{\'a}nchez}}, \ and\ \bibinfo {author} {\bibfnamefont
  {W.}~\bibnamefont {Yang}},\ }\href {\doibase 10.1126/science.1158722}
  {\bibfield  {journal} {\bibinfo  {journal} {Science}\ }\textbf {\bibinfo
  {volume} {321}},\ \bibinfo {pages} {792} (\bibinfo {year}
  {2008})}\BibitemShut {NoStop}%
\bibitem [{\citenamefont {Cohen}\ \emph {et~al.}(2012)\citenamefont {Cohen},
  \citenamefont {Mori-S{\'a}nchez},\ and\ \citenamefont
  {Yang}}]{Cohen-ChemRev-2012}%
  \BibitemOpen
  \bibfield  {author} {\bibinfo {author} {\bibfnamefont {A.~J.}\ \bibnamefont
  {Cohen}}, \bibinfo {author} {\bibfnamefont {P.}~\bibnamefont
  {Mori-S{\'a}nchez}}, \ and\ \bibinfo {author} {\bibfnamefont
  {W.}~\bibnamefont {Yang}},\ }\href {\doibase 10.1021/cr200107z} {\bibfield
  {journal} {\bibinfo  {journal} {Chemical Reviews}\ }\textbf {\bibinfo
  {volume} {112}},\ \bibinfo {pages} {289} (\bibinfo {year}
  {2012})}\BibitemShut {NoStop}%
\bibitem [{\citenamefont {Santra}\ and\ \citenamefont
  {Perdew}(2019)}]{Santra-JCP-2019}%
  \BibitemOpen
  \bibfield  {author} {\bibinfo {author} {\bibfnamefont {B.}~\bibnamefont
  {Santra}}\ and\ \bibinfo {author} {\bibfnamefont {J.~P.}\ \bibnamefont
  {Perdew}},\ }\href {\doibase 10.1063/1.5090534} {\bibfield  {journal}
  {\bibinfo  {journal} {The Journal of Chemical Physics}\ }\textbf {\bibinfo
  {volume} {150}},\ \bibinfo {pages} {174106} (\bibinfo {year}
  {2019})}\BibitemShut {NoStop}%
\bibitem [{\citenamefont {Zunger}\ \emph {et~al.}(2010)\citenamefont {Zunger},
  \citenamefont {Lany},\ and\ \citenamefont {Raebiger}}]{Zunger-Physics-2010}%
  \BibitemOpen
  \bibfield  {author} {\bibinfo {author} {\bibfnamefont {A.}~\bibnamefont
  {Zunger}}, \bibinfo {author} {\bibfnamefont {S.}~\bibnamefont {Lany}}, \ and\
  \bibinfo {author} {\bibfnamefont {H.}~\bibnamefont {Raebiger}},\ }\href
  {\doibase 10.1103/Physics.3.53} {\bibfield  {journal} {\bibinfo  {journal}
  {Physics}\ }\textbf {\bibinfo {volume} {3}},\ \bibinfo {pages} {53} (\bibinfo
  {year} {2010})}\BibitemShut {NoStop}%
\bibitem [{\citenamefont {Adeagbo}\ \emph {et~al.}(2014)\citenamefont
  {Adeagbo}, \citenamefont {Thomas}, \citenamefont {Nayak}, \citenamefont
  {Ernst},\ and\ \citenamefont {Hergert}}]{Adeagbo-PRB-2014}%
  \BibitemOpen
  \bibfield  {author} {\bibinfo {author} {\bibfnamefont {W.~A.}\ \bibnamefont
  {Adeagbo}}, \bibinfo {author} {\bibfnamefont {S.}~\bibnamefont {Thomas}},
  \bibinfo {author} {\bibfnamefont {S.~K.}\ \bibnamefont {Nayak}}, \bibinfo
  {author} {\bibfnamefont {A.}~\bibnamefont {Ernst}}, \ and\ \bibinfo {author}
  {\bibfnamefont {W.}~\bibnamefont {Hergert}},\ }\href {\doibase
  10.1103/PhysRevB.89.195135} {\bibfield  {journal} {\bibinfo  {journal} {Phys.
  Rev. B}\ }\textbf {\bibinfo {volume} {89}},\ \bibinfo {pages} {195135}
  (\bibinfo {year} {2014})}\BibitemShut {NoStop}%
\bibitem [{\citenamefont {Nayak}\ \emph {et~al.}(2015)\citenamefont {Nayak},
  \citenamefont {Langhammer}, \citenamefont {Adeagbo}, \citenamefont {Hergert},
  \citenamefont {M\"uller},\ and\ \citenamefont {B\"ottcher}}]{Nayak-PRB-2015}%
  \BibitemOpen
  \bibfield  {author} {\bibinfo {author} {\bibfnamefont {S.~K.}\ \bibnamefont
  {Nayak}}, \bibinfo {author} {\bibfnamefont {H.~T.}\ \bibnamefont
  {Langhammer}}, \bibinfo {author} {\bibfnamefont {W.~A.}\ \bibnamefont
  {Adeagbo}}, \bibinfo {author} {\bibfnamefont {W.}~\bibnamefont {Hergert}},
  \bibinfo {author} {\bibfnamefont {T.}~\bibnamefont {M\"uller}}, \ and\
  \bibinfo {author} {\bibfnamefont {R.}~\bibnamefont {B\"ottcher}},\ }\href
  {\doibase 10.1103/PhysRevB.91.155105} {\bibfield  {journal} {\bibinfo
  {journal} {Phys. Rev. B}\ }\textbf {\bibinfo {volume} {91}},\ \bibinfo
  {pages} {155105} (\bibinfo {year} {2015})}\BibitemShut {NoStop}%
\bibitem [{\citenamefont {Dion}\ \emph {et~al.}(2004)\citenamefont {Dion},
  \citenamefont {Rydberg}, \citenamefont {Schr\"oder}, \citenamefont
  {Langreth},\ and\ \citenamefont {Lundqvist}}]{Dion-PRL-2004}%
  \BibitemOpen
  \bibfield  {author} {\bibinfo {author} {\bibfnamefont {M.}~\bibnamefont
  {Dion}}, \bibinfo {author} {\bibfnamefont {H.}~\bibnamefont {Rydberg}},
  \bibinfo {author} {\bibfnamefont {E.}~\bibnamefont {Schr\"oder}}, \bibinfo
  {author} {\bibfnamefont {D.~C.}\ \bibnamefont {Langreth}}, \ and\ \bibinfo
  {author} {\bibfnamefont {B.~I.}\ \bibnamefont {Lundqvist}},\ }\href {\doibase
  10.1103/PhysRevLett.92.246401} {\bibfield  {journal} {\bibinfo  {journal}
  {Phys. Rev. Lett.}\ }\textbf {\bibinfo {volume} {92}},\ \bibinfo {pages}
  {246401} (\bibinfo {year} {2004})}\BibitemShut {NoStop}%
\bibitem [{\citenamefont {Berland}\ \emph {et~al.}(2019)\citenamefont
  {Berland}, \citenamefont {Chakraborty},\ and\ \citenamefont
  {Thonhauser}}]{Berland-PRB-2019}%
  \BibitemOpen
  \bibfield  {author} {\bibinfo {author} {\bibfnamefont {K.}~\bibnamefont
  {Berland}}, \bibinfo {author} {\bibfnamefont {D.}~\bibnamefont
  {Chakraborty}}, \ and\ \bibinfo {author} {\bibfnamefont {T.}~\bibnamefont
  {Thonhauser}},\ }\href {\doibase 10.1103/PhysRevB.99.195418} {\bibfield
  {journal} {\bibinfo  {journal} {Phys. Rev. B}\ }\textbf {\bibinfo {volume}
  {99}},\ \bibinfo {pages} {195418} (\bibinfo {year} {2019})}\BibitemShut
  {NoStop}%
\bibitem [{\citenamefont {Liechtenstein}\ \emph {et~al.}(1995)\citenamefont
  {Liechtenstein}, \citenamefont {Anisimov},\ and\ \citenamefont
  {Zaanen}}]{Liechtenstein-PRB-1995}%
  \BibitemOpen
  \bibfield  {author} {\bibinfo {author} {\bibfnamefont {A.~I.}\ \bibnamefont
  {Liechtenstein}}, \bibinfo {author} {\bibfnamefont {V.~I.}\ \bibnamefont
  {Anisimov}}, \ and\ \bibinfo {author} {\bibfnamefont {J.}~\bibnamefont
  {Zaanen}},\ }\href {\doibase 10.1103/PhysRevB.52.R5467} {\bibfield  {journal}
  {\bibinfo  {journal} {Phys. Rev. B}\ }\textbf {\bibinfo {volume} {52}},\
  \bibinfo {pages} {R5467} (\bibinfo {year} {1995})}\BibitemShut {NoStop}%
\bibitem [{\citenamefont {Becke}\ and\ \citenamefont
  {Johnson}(2006)}]{Becke-JCP-2006}%
  \BibitemOpen
  \bibfield  {author} {\bibinfo {author} {\bibfnamefont {A.~D.}\ \bibnamefont
  {Becke}}\ and\ \bibinfo {author} {\bibfnamefont {E.~R.}\ \bibnamefont
  {Johnson}},\ }\href {\doibase 10.1063/1.2213970} {\bibfield  {journal}
  {\bibinfo  {journal} {The Journal of Chemical Physics}\ }\textbf {\bibinfo
  {volume} {124}},\ \bibinfo {pages} {221101} (\bibinfo {year}
  {2006})}\BibitemShut {NoStop}%
\bibitem [{\citenamefont {Zhao}\ and\ \citenamefont
  {Truhlar}(2006)}]{Zhao-JCP-2006}%
  \BibitemOpen
  \bibfield  {author} {\bibinfo {author} {\bibfnamefont {Y.}~\bibnamefont
  {Zhao}}\ and\ \bibinfo {author} {\bibfnamefont {D.~G.}\ \bibnamefont
  {Truhlar}},\ }\href {\doibase 10.1063/1.2370993} {\bibfield  {journal}
  {\bibinfo  {journal} {The Journal of Chemical Physics}\ }\textbf {\bibinfo
  {volume} {125}},\ \bibinfo {pages} {194101} (\bibinfo {year}
  {2006})}\BibitemShut {NoStop}%
\bibitem [{\citenamefont {Yazyev}\ \emph {et~al.}(2012)\citenamefont {Yazyev},
  \citenamefont {Kioupakis}, \citenamefont {Moore},\ and\ \citenamefont
  {Louie}}]{Yazyev-PRB-2012}%
  \BibitemOpen
  \bibfield  {author} {\bibinfo {author} {\bibfnamefont {O.~V.}\ \bibnamefont
  {Yazyev}}, \bibinfo {author} {\bibfnamefont {E.}~\bibnamefont {Kioupakis}},
  \bibinfo {author} {\bibfnamefont {J.~E.}\ \bibnamefont {Moore}}, \ and\
  \bibinfo {author} {\bibfnamefont {S.~G.}\ \bibnamefont {Louie}},\ }\href
  {\doibase 10.1103/PhysRevB.85.161101} {\bibfield  {journal} {\bibinfo
  {journal} {Phys. Rev. B}\ }\textbf {\bibinfo {volume} {85}},\ \bibinfo
  {pages} {161101} (\bibinfo {year} {2012})}\BibitemShut {NoStop}%
\bibitem [{\citenamefont {Yazyev}\ \emph {et~al.}(2010)\citenamefont {Yazyev},
  \citenamefont {Moore},\ and\ \citenamefont {Louie}}]{Yazyev-PRL-2010}%
  \BibitemOpen
  \bibfield  {author} {\bibinfo {author} {\bibfnamefont {O.~V.}\ \bibnamefont
  {Yazyev}}, \bibinfo {author} {\bibfnamefont {J.~E.}\ \bibnamefont {Moore}}, \
  and\ \bibinfo {author} {\bibfnamefont {S.~G.}\ \bibnamefont {Louie}},\ }\href
  {\doibase 10.1103/PhysRevLett.105.266806} {\bibfield  {journal} {\bibinfo
  {journal} {Phys. Rev. Lett.}\ }\textbf {\bibinfo {volume} {105}},\ \bibinfo
  {pages} {266806} (\bibinfo {year} {2010})}\BibitemShut {NoStop}%
\bibitem [{\citenamefont {Li}\ \emph {et~al.}(2014)\citenamefont {Li},
  \citenamefont {Wei}, \citenamefont {Zhu}, \citenamefont {Ting},\ and\
  \citenamefont {Chen}}]{Li-PRB-2014}%
  \BibitemOpen
  \bibfield  {author} {\bibinfo {author} {\bibfnamefont {W.}~\bibnamefont
  {Li}}, \bibinfo {author} {\bibfnamefont {X.-Y.}\ \bibnamefont {Wei}},
  \bibinfo {author} {\bibfnamefont {J.-X.}\ \bibnamefont {Zhu}}, \bibinfo
  {author} {\bibfnamefont {C.~S.}\ \bibnamefont {Ting}}, \ and\ \bibinfo
  {author} {\bibfnamefont {Y.}~\bibnamefont {Chen}},\ }\href {\doibase
  10.1103/PhysRevB.89.035101} {\bibfield  {journal} {\bibinfo  {journal} {Phys.
  Rev. B}\ }\textbf {\bibinfo {volume} {89}},\ \bibinfo {pages} {035101}
  (\bibinfo {year} {2014})}\BibitemShut {NoStop}%
\bibitem [{\citenamefont {Tao}\ and\ \citenamefont
  {Tsymbal}(2018)}]{Tao-NP-2018}%
  \BibitemOpen
  \bibfield  {author} {\bibinfo {author} {\bibfnamefont {L.~L.}\ \bibnamefont
  {Tao}}\ and\ \bibinfo {author} {\bibfnamefont {E.~Y.}\ \bibnamefont
  {Tsymbal}},\ }\href {\doibase https://10.1038/s41467-018-05137-0} {\bibfield
  {journal} {\bibinfo  {journal} {Nature Communications}\ }\textbf {\bibinfo
  {volume} {9}},\ \bibinfo {pages} {2763} (\bibinfo {year} {2018})}\BibitemShut
  {NoStop}%
\bibitem [{\citenamefont {Datzer}\ \emph {et~al.}(2017)\citenamefont {Datzer},
  \citenamefont {Zumb\"ulte}, \citenamefont {Braun}, \citenamefont {F\"orster},
  \citenamefont {Schmidt}, \citenamefont {Mi}, \citenamefont {Iversen},
  \citenamefont {Hofmann}, \citenamefont {Min\'ar}, \citenamefont {Ebert},
  \citenamefont {Kr\"uger}, \citenamefont {Rohlfing},\ and\ \citenamefont
  {Donath}}]{Datzer-PRB-2017}%
  \BibitemOpen
  \bibfield  {author} {\bibinfo {author} {\bibfnamefont {C.}~\bibnamefont
  {Datzer}}, \bibinfo {author} {\bibfnamefont {A.}~\bibnamefont {Zumb\"ulte}},
  \bibinfo {author} {\bibfnamefont {J.}~\bibnamefont {Braun}}, \bibinfo
  {author} {\bibfnamefont {T.}~\bibnamefont {F\"orster}}, \bibinfo {author}
  {\bibfnamefont {A.~B.}\ \bibnamefont {Schmidt}}, \bibinfo {author}
  {\bibfnamefont {J.}~\bibnamefont {Mi}}, \bibinfo {author} {\bibfnamefont
  {B.}~\bibnamefont {Iversen}}, \bibinfo {author} {\bibfnamefont
  {P.}~\bibnamefont {Hofmann}}, \bibinfo {author} {\bibfnamefont
  {J.}~\bibnamefont {Min\'ar}}, \bibinfo {author} {\bibfnamefont
  {H.}~\bibnamefont {Ebert}}, \bibinfo {author} {\bibfnamefont
  {P.}~\bibnamefont {Kr\"uger}}, \bibinfo {author} {\bibfnamefont
  {M.}~\bibnamefont {Rohlfing}}, \ and\ \bibinfo {author} {\bibfnamefont
  {M.}~\bibnamefont {Donath}},\ }\href {\doibase 10.1103/PhysRevB.95.115401}
  {\bibfield  {journal} {\bibinfo  {journal} {Phys. Rev. B}\ }\textbf {\bibinfo
  {volume} {95}},\ \bibinfo {pages} {115401} (\bibinfo {year}
  {2017})}\BibitemShut {NoStop}%
\bibitem [{\citenamefont {Seixas}\ \emph {et~al.}(2015)\citenamefont {Seixas},
  \citenamefont {West}, \citenamefont {Fazzio},\ and\ \citenamefont
  {Zhang}}]{Seixas-NatureComm-2015}%
  \BibitemOpen
  \bibfield  {author} {\bibinfo {author} {\bibfnamefont {L.}~\bibnamefont
  {Seixas}}, \bibinfo {author} {\bibfnamefont {D.}~\bibnamefont {West}},
  \bibinfo {author} {\bibfnamefont {A.}~\bibnamefont {Fazzio}}, \ and\ \bibinfo
  {author} {\bibfnamefont {S.~B.}\ \bibnamefont {Zhang}},\ }\href {\doibase
  https://0.1038/ncomms8630} {\bibfield  {journal} {\bibinfo  {journal} {Nature
  Communications}\ }\textbf {\bibinfo {volume} {6}},\ \bibinfo {pages} {7360}
  (\bibinfo {year} {2015})}\BibitemShut {NoStop}%
\bibitem [{\citenamefont {Seibel}\ \emph {et~al.}(2016)\citenamefont {Seibel},
  \citenamefont {Braun}, \citenamefont {Maa\ss{}}, \citenamefont {Bentmann},
  \citenamefont {Min\'ar}, \citenamefont {Kuznetsova}, \citenamefont {Kokh},
  \citenamefont {Tereshchenko}, \citenamefont {Okuda}, \citenamefont {Ebert},\
  and\ \citenamefont {Reinert}}]{Seibel-PRB-2016}%
  \BibitemOpen
  \bibfield  {author} {\bibinfo {author} {\bibfnamefont {C.}~\bibnamefont
  {Seibel}}, \bibinfo {author} {\bibfnamefont {J.}~\bibnamefont {Braun}},
  \bibinfo {author} {\bibfnamefont {H.}~\bibnamefont {Maa\ss{}}}, \bibinfo
  {author} {\bibfnamefont {H.}~\bibnamefont {Bentmann}}, \bibinfo {author}
  {\bibfnamefont {J.}~\bibnamefont {Min\'ar}}, \bibinfo {author} {\bibfnamefont
  {T.~V.}\ \bibnamefont {Kuznetsova}}, \bibinfo {author} {\bibfnamefont
  {K.~A.}\ \bibnamefont {Kokh}}, \bibinfo {author} {\bibfnamefont {O.~E.}\
  \bibnamefont {Tereshchenko}}, \bibinfo {author} {\bibfnamefont
  {T.}~\bibnamefont {Okuda}}, \bibinfo {author} {\bibfnamefont
  {H.}~\bibnamefont {Ebert}}, \ and\ \bibinfo {author} {\bibfnamefont
  {F.}~\bibnamefont {Reinert}},\ }\href {\doibase 10.1103/PhysRevB.93.245150}
  {\bibfield  {journal} {\bibinfo  {journal} {Phys. Rev. B}\ }\textbf {\bibinfo
  {volume} {93}},\ \bibinfo {pages} {245150} (\bibinfo {year}
  {2016})}\BibitemShut {NoStop}%
\bibitem [{\citenamefont {Rusinov}\ \emph {et~al.}(2013)\citenamefont
  {Rusinov}, \citenamefont {Nechaev},\ and\ \citenamefont
  {Chulkov}}]{Rusinov-JExptTheoPhys-2013}%
  \BibitemOpen
  \bibfield  {author} {\bibinfo {author} {\bibfnamefont {I.~P.}\ \bibnamefont
  {Rusinov}}, \bibinfo {author} {\bibfnamefont {I.~A.}\ \bibnamefont
  {Nechaev}}, \ and\ \bibinfo {author} {\bibfnamefont {E.~V.}\ \bibnamefont
  {Chulkov}},\ }\href {\doibase 10.1134/S1063776113060216} {\bibfield
  {journal} {\bibinfo  {journal} {Journal of Experimental and Theoretical
  Physics}\ }\textbf {\bibinfo {volume} {116}},\ \bibinfo {pages} {1006}
  (\bibinfo {year} {2013})}\BibitemShut {NoStop}%
\bibitem [{\citenamefont {Aguilera}\ \emph {et~al.}(2013)\citenamefont
  {Aguilera}, \citenamefont {Friedrich}, \citenamefont {Bihlmayer},\ and\
  \citenamefont {Bl\"ugel}}]{Aguilera-PRB-2013}%
  \BibitemOpen
  \bibfield  {author} {\bibinfo {author} {\bibfnamefont {I.}~\bibnamefont
  {Aguilera}}, \bibinfo {author} {\bibfnamefont {C.}~\bibnamefont {Friedrich}},
  \bibinfo {author} {\bibfnamefont {G.}~\bibnamefont {Bihlmayer}}, \ and\
  \bibinfo {author} {\bibfnamefont {S.}~\bibnamefont {Bl\"ugel}},\ }\href
  {\doibase 10.1103/PhysRevB.88.045206} {\bibfield  {journal} {\bibinfo
  {journal} {Phys. Rev. B}\ }\textbf {\bibinfo {volume} {88}},\ \bibinfo
  {pages} {045206} (\bibinfo {year} {2013})}\BibitemShut {NoStop}%
\bibitem [{\citenamefont {F\"orster}\ \emph {et~al.}(2015)\citenamefont
  {F\"orster}, \citenamefont {Kr\"uger},\ and\ \citenamefont
  {Rohlfing}}]{Forster-PRB-2015}%
  \BibitemOpen
  \bibfield  {author} {\bibinfo {author} {\bibfnamefont {T.}~\bibnamefont
  {F\"orster}}, \bibinfo {author} {\bibfnamefont {P.}~\bibnamefont {Kr\"uger}},
  \ and\ \bibinfo {author} {\bibfnamefont {M.}~\bibnamefont {Rohlfing}},\
  }\href {\doibase 10.1103/PhysRevB.92.201404} {\bibfield  {journal} {\bibinfo
  {journal} {Phys. Rev. B}\ }\textbf {\bibinfo {volume} {92}},\ \bibinfo
  {pages} {201404} (\bibinfo {year} {2015})}\BibitemShut {NoStop}%
\bibitem [{\citenamefont {F\"orster}\ \emph {et~al.}(2016)\citenamefont
  {F\"orster}, \citenamefont {Kr\"uger},\ and\ \citenamefont
  {Rohlfing}}]{Forster-PRB-2016}%
  \BibitemOpen
  \bibfield  {author} {\bibinfo {author} {\bibfnamefont {T.}~\bibnamefont
  {F\"orster}}, \bibinfo {author} {\bibfnamefont {P.}~\bibnamefont {Kr\"uger}},
  \ and\ \bibinfo {author} {\bibfnamefont {M.}~\bibnamefont {Rohlfing}},\
  }\href {\doibase 10.1103/PhysRevB.93.205442} {\bibfield  {journal} {\bibinfo
  {journal} {Phys. Rev. B}\ }\textbf {\bibinfo {volume} {93}},\ \bibinfo
  {pages} {205442} (\bibinfo {year} {2016})}\BibitemShut {NoStop}%
\bibitem [{\citenamefont {Liu}\ \emph {et~al.}(2013)\citenamefont {Liu},
  \citenamefont {Peng}, \citenamefont {Wei}, \citenamefont {Yang},
  \citenamefont {Stocks},\ and\ \citenamefont {Zhong}}]{Liu-PRB-2013}%
  \BibitemOpen
  \bibfield  {author} {\bibinfo {author} {\bibfnamefont {W.}~\bibnamefont
  {Liu}}, \bibinfo {author} {\bibfnamefont {X.}~\bibnamefont {Peng}}, \bibinfo
  {author} {\bibfnamefont {X.}~\bibnamefont {Wei}}, \bibinfo {author}
  {\bibfnamefont {H.}~\bibnamefont {Yang}}, \bibinfo {author} {\bibfnamefont
  {G.~M.}\ \bibnamefont {Stocks}}, \ and\ \bibinfo {author} {\bibfnamefont
  {J.}~\bibnamefont {Zhong}},\ }\href {\doibase 10.1103/PhysRevB.87.205315}
  {\bibfield  {journal} {\bibinfo  {journal} {Phys. Rev. B}\ }\textbf {\bibinfo
  {volume} {87}},\ \bibinfo {pages} {205315} (\bibinfo {year}
  {2013})}\BibitemShut {NoStop}%
\bibitem [{\citenamefont {Luo}\ \emph {et~al.}(2012)\citenamefont {Luo},
  \citenamefont {Sullivan},\ and\ \citenamefont {Quek}}]{Luo-PRB-2012}%
  \BibitemOpen
  \bibfield  {author} {\bibinfo {author} {\bibfnamefont {X.}~\bibnamefont
  {Luo}}, \bibinfo {author} {\bibfnamefont {M.~B.}\ \bibnamefont {Sullivan}}, \
  and\ \bibinfo {author} {\bibfnamefont {S.~Y.}\ \bibnamefont {Quek}},\ }\href
  {\doibase 10.1103/PhysRevB.86.184111} {\bibfield  {journal} {\bibinfo
  {journal} {Phys. Rev. B}\ }\textbf {\bibinfo {volume} {86}},\ \bibinfo
  {pages} {184111} (\bibinfo {year} {2012})}\BibitemShut {NoStop}%
\bibitem [{\citenamefont {Grimme}\ \emph {et~al.}(2010)\citenamefont {Grimme},
  \citenamefont {Antony}, \citenamefont {Ehrlich},\ and\ \citenamefont
  {Krieg}}]{Grimme-JCP-2010}%
  \BibitemOpen
  \bibfield  {author} {\bibinfo {author} {\bibfnamefont {S.}~\bibnamefont
  {Grimme}}, \bibinfo {author} {\bibfnamefont {J.}~\bibnamefont {Antony}},
  \bibinfo {author} {\bibfnamefont {S.}~\bibnamefont {Ehrlich}}, \ and\
  \bibinfo {author} {\bibfnamefont {H.}~\bibnamefont {Krieg}},\ }\href
  {\doibase 10.1063/1.3382344} {\bibfield  {journal} {\bibinfo  {journal} {The
  Journal of Chemical Physics}\ }\textbf {\bibinfo {volume} {132}},\ \bibinfo
  {pages} {154104} (\bibinfo {year} {2010})}\BibitemShut {NoStop}%
\bibitem [{\citenamefont {Kresse}\ and\ \citenamefont
  {Furthm\"{u}ller}(1996{\natexlab{a}})}]{Kresse-CMS-1996}%
  \BibitemOpen
  \bibfield  {author} {\bibinfo {author} {\bibfnamefont {G.}~\bibnamefont
  {Kresse}}\ and\ \bibinfo {author} {\bibfnamefont {J.}~\bibnamefont
  {Furthm\"{u}ller}},\ }\href {\doibase
  http://dx.doi.org/10.1016/0927-0256(96)00008-0} {\bibfield  {journal}
  {\bibinfo  {journal} {Computational Materials Science}\ }\textbf {\bibinfo
  {volume} {6}},\ \bibinfo {pages} {15} (\bibinfo {year}
  {1996}{\natexlab{a}})}\BibitemShut {NoStop}%
\bibitem [{\citenamefont {Kresse}\ and\ \citenamefont
  {Furthm\"{u}ller}(1996{\natexlab{b}})}]{Kresse-PRB-1996}%
  \BibitemOpen
  \bibfield  {author} {\bibinfo {author} {\bibfnamefont {G.}~\bibnamefont
  {Kresse}}\ and\ \bibinfo {author} {\bibfnamefont {J.}~\bibnamefont
  {Furthm\"{u}ller}},\ }\href {\doibase 10.1103/PhysRevB.54.11169} {\bibfield
  {journal} {\bibinfo  {journal} {Phys. Rev. B}\ }\textbf {\bibinfo {volume}
  {54}},\ \bibinfo {pages} {11169} (\bibinfo {year}
  {1996}{\natexlab{b}})}\BibitemShut {NoStop}%
\bibitem [{\citenamefont {Ceperley}\ and\ \citenamefont
  {Alder}(1980)}]{Ceperley-PRL-1980}%
  \BibitemOpen
  \bibfield  {author} {\bibinfo {author} {\bibfnamefont {D.~M.}\ \bibnamefont
  {Ceperley}}\ and\ \bibinfo {author} {\bibfnamefont {B.~J.}\ \bibnamefont
  {Alder}},\ }\href {\doibase 10.1103/PhysRevLett.45.566} {\bibfield  {journal}
  {\bibinfo  {journal} {Phys. Rev. Lett.}\ }\textbf {\bibinfo {volume} {45}},\
  \bibinfo {pages} {566} (\bibinfo {year} {1980})}\BibitemShut {NoStop}%
\bibitem [{\citenamefont {Perdew}\ \emph {et~al.}(1996)\citenamefont {Perdew},
  \citenamefont {Burke},\ and\ \citenamefont {Ernzerhof}}]{Perdew-PRL-1996}%
  \BibitemOpen
  \bibfield  {author} {\bibinfo {author} {\bibfnamefont {J.~P.}\ \bibnamefont
  {Perdew}}, \bibinfo {author} {\bibfnamefont {K.}~\bibnamefont {Burke}}, \
  and\ \bibinfo {author} {\bibfnamefont {M.}~\bibnamefont {Ernzerhof}},\ }\href
  {\doibase 10.1103/PhysRevLett.77.3865} {\bibfield  {journal} {\bibinfo
  {journal} {Phys. Rev. Lett.}\ }\textbf {\bibinfo {volume} {77}},\ \bibinfo
  {pages} {3865} (\bibinfo {year} {1996})}\BibitemShut {NoStop}%
\bibitem [{\citenamefont {Sun}\ \emph {et~al.}(2015)\citenamefont {Sun},
  \citenamefont {Ruzsinszky},\ and\ \citenamefont {Perdew}}]{Sun-PRL-2015}%
  \BibitemOpen
  \bibfield  {author} {\bibinfo {author} {\bibfnamefont {J.}~\bibnamefont
  {Sun}}, \bibinfo {author} {\bibfnamefont {A.}~\bibnamefont {Ruzsinszky}}, \
  and\ \bibinfo {author} {\bibfnamefont {J.~P.}\ \bibnamefont {Perdew}},\
  }\href {\doibase 10.1103/PhysRevLett.115.036402} {\bibfield  {journal}
  {\bibinfo  {journal} {Phys. Rev. Lett.}\ }\textbf {\bibinfo {volume} {115}},\
  \bibinfo {pages} {036402} (\bibinfo {year} {2015})}\BibitemShut {NoStop}%
\bibitem [{\citenamefont {Sahoo}\ \emph {et~al.}(2018)\citenamefont {Sahoo},
  \citenamefont {Alpay},\ and\ \citenamefont
  {Hebert}}]{Sahoo-SurfaceScience-2018}%
  \BibitemOpen
  \bibfield  {author} {\bibinfo {author} {\bibfnamefont {S.}~\bibnamefont
  {Sahoo}}, \bibinfo {author} {\bibfnamefont {S.~P.}\ \bibnamefont {Alpay}}, \
  and\ \bibinfo {author} {\bibfnamefont {R.~J.}\ \bibnamefont {Hebert}},\
  }\href {\doibase https://doi.org/10.1016/j.susc.2018.05.007} {\bibfield
  {journal} {\bibinfo  {journal} {Surface Science}\ }\textbf {\bibinfo {volume}
  {677}},\ \bibinfo {pages} {18 } (\bibinfo {year} {2018})}\BibitemShut
  {NoStop}%
\bibitem [{\citenamefont {Sahoo}\ \emph {et~al.}(2010)\citenamefont {Sahoo},
  \citenamefont {Hucht}, \citenamefont {Gruner}, \citenamefont {Rollmann},
  \citenamefont {Entel}, \citenamefont {Postnikov}, \citenamefont {Ferrer},
  \citenamefont {Fern\'andez-Seivane}, \citenamefont {Richter}, \citenamefont
  {Fritsch},\ and\ \citenamefont {Sil}}]{Sahoo-PRB-2010}%
  \BibitemOpen
  \bibfield  {author} {\bibinfo {author} {\bibfnamefont {S.}~\bibnamefont
  {Sahoo}}, \bibinfo {author} {\bibfnamefont {A.}~\bibnamefont {Hucht}},
  \bibinfo {author} {\bibfnamefont {M.~E.}\ \bibnamefont {Gruner}}, \bibinfo
  {author} {\bibfnamefont {G.}~\bibnamefont {Rollmann}}, \bibinfo {author}
  {\bibfnamefont {P.}~\bibnamefont {Entel}}, \bibinfo {author} {\bibfnamefont
  {A.}~\bibnamefont {Postnikov}}, \bibinfo {author} {\bibfnamefont
  {J.}~\bibnamefont {Ferrer}}, \bibinfo {author} {\bibfnamefont
  {L.}~\bibnamefont {Fern\'andez-Seivane}}, \bibinfo {author} {\bibfnamefont
  {M.}~\bibnamefont {Richter}}, \bibinfo {author} {\bibfnamefont
  {D.}~\bibnamefont {Fritsch}}, \ and\ \bibinfo {author} {\bibfnamefont
  {S.}~\bibnamefont {Sil}},\ }\href {\doibase 10.1103/PhysRevB.82.054418}
  {\bibfield  {journal} {\bibinfo  {journal} {Phys. Rev. B}\ }\textbf {\bibinfo
  {volume} {82}},\ \bibinfo {pages} {054418} (\bibinfo {year}
  {2010})}\BibitemShut {NoStop}%
\bibitem [{\citenamefont {Martinez}\ \emph {et~al.}(2017)\citenamefont
  {Martinez}, \citenamefont {Piot}, \citenamefont {Potemski}, \citenamefont
  {Hor}, \citenamefont {Materna}, \citenamefont {Strzelecka}, \citenamefont
  {Hruban}, \citenamefont {Caha}, \citenamefont {Nov\'{a}k}, \citenamefont
  {Dubroka}, \citenamefont {Dra\v{s}ar},\ and\ \citenamefont
  {Orlita}}]{Martinez-SciRep-2017}%
  \BibitemOpen
  \bibfield  {author} {\bibinfo {author} {\bibfnamefont {G.}~\bibnamefont
  {Martinez}}, \bibinfo {author} {\bibfnamefont {B.~A.}\ \bibnamefont {Piot}},
  \bibinfo {author} {\bibfnamefont {M.}~\bibnamefont {Potemski}}, \bibinfo
  {author} {\bibfnamefont {Y.~S.}\ \bibnamefont {Hor}}, \bibinfo {author}
  {\bibfnamefont {A.}~\bibnamefont {Materna}}, \bibinfo {author} {\bibfnamefont
  {S.~G.}\ \bibnamefont {Strzelecka}}, \bibinfo {author} {\bibfnamefont
  {A.}~\bibnamefont {Hruban}}, \bibinfo {author} {\bibfnamefont
  {O.}~\bibnamefont {Caha}}, \bibinfo {author} {\bibfnamefont {J.}~\bibnamefont
  {Nov\'{a}k}}, \bibinfo {author} {\bibfnamefont {A.}~\bibnamefont {Dubroka}},
  \bibinfo {author} {\bibfnamefont {v.}~\bibnamefont {Dra\v{s}ar}}, \ and\
  \bibinfo {author} {\bibfnamefont {M.}~\bibnamefont {Orlita}},\ }\href
  {\doibase 10.1038/s41598-017-07211-x} {\bibfield  {journal} {\bibinfo
  {journal} {Scientific Reports}\ }\textbf {\bibinfo {volume} {7}},\ \bibinfo
  {pages} {6891} (\bibinfo {year} {2017})}\BibitemShut {NoStop}%
\bibitem [{\citenamefont {Nechaev}\ \emph {et~al.}(2013)\citenamefont
  {Nechaev}, \citenamefont {Hatch}, \citenamefont {Bianchi}, \citenamefont
  {Guan}, \citenamefont {Friedrich}, \citenamefont {Aguilera}, \citenamefont
  {Mi}, \citenamefont {Iversen}, \citenamefont {Bl\"ugel}, \citenamefont
  {Hofmann},\ and\ \citenamefont {Chulkov}}]{Nechaev-PRB-2013}%
  \BibitemOpen
  \bibfield  {author} {\bibinfo {author} {\bibfnamefont {I.~A.}\ \bibnamefont
  {Nechaev}}, \bibinfo {author} {\bibfnamefont {R.~C.}\ \bibnamefont {Hatch}},
  \bibinfo {author} {\bibfnamefont {M.}~\bibnamefont {Bianchi}}, \bibinfo
  {author} {\bibfnamefont {D.}~\bibnamefont {Guan}}, \bibinfo {author}
  {\bibfnamefont {C.}~\bibnamefont {Friedrich}}, \bibinfo {author}
  {\bibfnamefont {I.}~\bibnamefont {Aguilera}}, \bibinfo {author}
  {\bibfnamefont {J.~L.}\ \bibnamefont {Mi}}, \bibinfo {author} {\bibfnamefont
  {B.~B.}\ \bibnamefont {Iversen}}, \bibinfo {author} {\bibfnamefont
  {S.}~\bibnamefont {Bl\"ugel}}, \bibinfo {author} {\bibfnamefont
  {P.}~\bibnamefont {Hofmann}}, \ and\ \bibinfo {author} {\bibfnamefont
  {E.~V.}\ \bibnamefont {Chulkov}},\ }\href {\doibase
  10.1103/PhysRevB.87.121111} {\bibfield  {journal} {\bibinfo  {journal} {Phys.
  Rev. B}\ }\textbf {\bibinfo {volume} {87}},\ \bibinfo {pages} {121111}
  (\bibinfo {year} {2013})}\BibitemShut {NoStop}%
\bibitem [{\citenamefont {Li}\ \emph {et~al.}(1961)\citenamefont {Li},
  \citenamefont {Ruoff},\ and\ \citenamefont {Spencer}}]{Che-Yu-JAP-1961}%
  \BibitemOpen
  \bibfield  {author} {\bibinfo {author} {\bibfnamefont {C.-Y.}\ \bibnamefont
  {Li}}, \bibinfo {author} {\bibfnamefont {A.~L.}\ \bibnamefont {Ruoff}}, \
  and\ \bibinfo {author} {\bibfnamefont {C.~W.}\ \bibnamefont {Spencer}},\
  }\href {\doibase 10.1063/1.1728426} {\bibfield  {journal} {\bibinfo
  {journal} {J. Appl. Phys.}\ }\textbf {\bibinfo {volume} {32}},\ \bibinfo
  {pages} {1733} (\bibinfo {year} {1961})}\BibitemShut {NoStop}%
\bibitem [{\citenamefont {Roy}\ \emph {et~al.}(2014{\natexlab{a}})\citenamefont
  {Roy}, \citenamefont {Meyerheim}, \citenamefont {Mohseni}, \citenamefont
  {Ernst}, \citenamefont {Otrokov}, \citenamefont {Vergniory}, \citenamefont
  {Mussler}, \citenamefont {Kampmeier}, \citenamefont {Gr\"utzmacher},
  \citenamefont {Tusche}, \citenamefont {Schneider}, \citenamefont {Chulkov},\
  and\ \citenamefont {Kirschner}}]{Roy-PRB-2014}%
  \BibitemOpen
  \bibfield  {author} {\bibinfo {author} {\bibfnamefont {S.}~\bibnamefont
  {Roy}}, \bibinfo {author} {\bibfnamefont {H.~L.}\ \bibnamefont {Meyerheim}},
  \bibinfo {author} {\bibfnamefont {K.}~\bibnamefont {Mohseni}}, \bibinfo
  {author} {\bibfnamefont {A.}~\bibnamefont {Ernst}}, \bibinfo {author}
  {\bibfnamefont {M.~M.}\ \bibnamefont {Otrokov}}, \bibinfo {author}
  {\bibfnamefont {M.~G.}\ \bibnamefont {Vergniory}}, \bibinfo {author}
  {\bibfnamefont {G.}~\bibnamefont {Mussler}}, \bibinfo {author} {\bibfnamefont
  {J.}~\bibnamefont {Kampmeier}}, \bibinfo {author} {\bibfnamefont
  {D.}~\bibnamefont {Gr\"utzmacher}}, \bibinfo {author} {\bibfnamefont
  {C.}~\bibnamefont {Tusche}}, \bibinfo {author} {\bibfnamefont
  {J.}~\bibnamefont {Schneider}}, \bibinfo {author} {\bibfnamefont {E.~V.}\
  \bibnamefont {Chulkov}}, \ and\ \bibinfo {author} {\bibfnamefont
  {J.}~\bibnamefont {Kirschner}},\ }\href {\doibase 10.1103/PhysRevB.90.155456}
  {\bibfield  {journal} {\bibinfo  {journal} {Phys. Rev. B}\ }\textbf {\bibinfo
  {volume} {90}},\ \bibinfo {pages} {155456} (\bibinfo {year}
  {2014}{\natexlab{a}})}\BibitemShut {NoStop}%
\bibitem [{\citenamefont {Fukui}\ \emph {et~al.}(2012)\citenamefont {Fukui},
  \citenamefont {Hirahara}, \citenamefont {Shirasawa}, \citenamefont
  {Takahashi}, \citenamefont {Kobayashi},\ and\ \citenamefont
  {Hasegawa}}]{Fukui-PRB-2012}%
  \BibitemOpen
  \bibfield  {author} {\bibinfo {author} {\bibfnamefont {N.}~\bibnamefont
  {Fukui}}, \bibinfo {author} {\bibfnamefont {T.}~\bibnamefont {Hirahara}},
  \bibinfo {author} {\bibfnamefont {T.}~\bibnamefont {Shirasawa}}, \bibinfo
  {author} {\bibfnamefont {T.}~\bibnamefont {Takahashi}}, \bibinfo {author}
  {\bibfnamefont {K.}~\bibnamefont {Kobayashi}}, \ and\ \bibinfo {author}
  {\bibfnamefont {S.}~\bibnamefont {Hasegawa}},\ }\href {\doibase
  10.1103/PhysRevB.85.115426} {\bibfield  {journal} {\bibinfo  {journal} {Phys.
  Rev. B}\ }\textbf {\bibinfo {volume} {85}},\ \bibinfo {pages} {115426}
  (\bibinfo {year} {2012})}\BibitemShut {NoStop}%
\bibitem [{\citenamefont {Roy}\ \emph {et~al.}(2014{\natexlab{b}})\citenamefont
  {Roy}, \citenamefont {Meyerheim}, \citenamefont {Ernst}, \citenamefont
  {Mohseni}, \citenamefont {Tusche}, \citenamefont {Vergniory}, \citenamefont
  {Menshchikova}, \citenamefont {Otrokov}, \citenamefont {Ryabishchenkova},
  \citenamefont {Aliev}, \citenamefont {Babanly}, \citenamefont {Kokh},
  \citenamefont {Tereshchenko}, \citenamefont {Chulkov}, \citenamefont
  {Schneider},\ and\ \citenamefont {Kirschner}}]{Roy-PRL-2014}%
  \BibitemOpen
  \bibfield  {author} {\bibinfo {author} {\bibfnamefont {S.}~\bibnamefont
  {Roy}}, \bibinfo {author} {\bibfnamefont {H.~L.}\ \bibnamefont {Meyerheim}},
  \bibinfo {author} {\bibfnamefont {A.}~\bibnamefont {Ernst}}, \bibinfo
  {author} {\bibfnamefont {K.}~\bibnamefont {Mohseni}}, \bibinfo {author}
  {\bibfnamefont {C.}~\bibnamefont {Tusche}}, \bibinfo {author} {\bibfnamefont
  {M.~G.}\ \bibnamefont {Vergniory}}, \bibinfo {author} {\bibfnamefont {T.~V.}\
  \bibnamefont {Menshchikova}}, \bibinfo {author} {\bibfnamefont {M.~M.}\
  \bibnamefont {Otrokov}}, \bibinfo {author} {\bibfnamefont {A.~G.}\
  \bibnamefont {Ryabishchenkova}}, \bibinfo {author} {\bibfnamefont {Z.~S.}\
  \bibnamefont {Aliev}}, \bibinfo {author} {\bibfnamefont {M.~B.}\ \bibnamefont
  {Babanly}}, \bibinfo {author} {\bibfnamefont {K.~A.}\ \bibnamefont {Kokh}},
  \bibinfo {author} {\bibfnamefont {O.~E.}\ \bibnamefont {Tereshchenko}},
  \bibinfo {author} {\bibfnamefont {E.~V.}\ \bibnamefont {Chulkov}}, \bibinfo
  {author} {\bibfnamefont {J.}~\bibnamefont {Schneider}}, \ and\ \bibinfo
  {author} {\bibfnamefont {J.}~\bibnamefont {Kirschner}},\ }\href {\doibase
  10.1103/PhysRevLett.113.116802} {\bibfield  {journal} {\bibinfo  {journal}
  {Phys. Rev. Lett.}\ }\textbf {\bibinfo {volume} {113}},\ \bibinfo {pages}
  {116802} (\bibinfo {year} {2014}{\natexlab{b}})}\BibitemShut {NoStop}%
\bibitem [{\citenamefont {Geilhufe}\ \emph {et~al.}(2015)\citenamefont
  {Geilhufe}, \citenamefont {Nayak}, \citenamefont {Thomas}, \citenamefont
  {D\"ane}, \citenamefont {Tripathi}, \citenamefont {Entel}, \citenamefont
  {Hergert},\ and\ \citenamefont {Ernst}}]{Geilhufe-PRB-2015}%
  \BibitemOpen
  \bibfield  {author} {\bibinfo {author} {\bibfnamefont {M.}~\bibnamefont
  {Geilhufe}}, \bibinfo {author} {\bibfnamefont {S.~K.}\ \bibnamefont {Nayak}},
  \bibinfo {author} {\bibfnamefont {S.}~\bibnamefont {Thomas}}, \bibinfo
  {author} {\bibfnamefont {M.}~\bibnamefont {D\"ane}}, \bibinfo {author}
  {\bibfnamefont {G.~S.}\ \bibnamefont {Tripathi}}, \bibinfo {author}
  {\bibfnamefont {P.}~\bibnamefont {Entel}}, \bibinfo {author} {\bibfnamefont
  {W.}~\bibnamefont {Hergert}}, \ and\ \bibinfo {author} {\bibfnamefont
  {A.}~\bibnamefont {Ernst}},\ }\href {\doibase 10.1103/PhysRevB.92.235203}
  {\bibfield  {journal} {\bibinfo  {journal} {Phys. Rev. B}\ }\textbf {\bibinfo
  {volume} {92}},\ \bibinfo {pages} {235203} (\bibinfo {year}
  {2015})}\BibitemShut {NoStop}%
\end{thebibliography}%
\end{document}